\documentclass[fleqn, usenatbib]{mnras}

\usepackage{newtxtext,newtxmath}

    \usepackage[T1]{fontenc}
\usepackage{ae,aecompl}

\def\Pdd{P_{\rm{\delta\delta}}}
\def\Pdt{P_{\rm{\delta\theta}}}
\def\Ptt{P_{\rm{\theta\theta}}}

\def\Pgg{P_{\rm{gg}}}
\def\Pgt{P_{\rm{g\theta}}}
\def\d{\mathrm d}

\usepackage{graphicx}	
\usepackage{amsmath}	

\usepackage{bigints}

\usepackage[ruled]{algorithm2e}

\usepackage{multirow}

\usepackage{multicol}

\usepackage{cancel}

\usepackage{siunitx}

\usepackage{diagbox}

\newcommand{\de}{\mathrm{d}}
\newcommand{\Mpch}{h^{-1} \, \mathrm{Mpc}}

\newcommand{\Leg}[2]{\mathcal{P}_{#1}(#2)}
\newcommand{\BarLeg}[2]{\overline{\mathcal{P}}_{#1}(#2)}

\renewcommand{\vec}[1]{\mathbf{#1}}

\title[Joint 2- and 3-point clustering analysis]
{A joint 2- and 3-point clustering analysis of the VIPERS PDR2 catalogue at $z\sim1$: breaking the degeneracy of cosmological parameters}

\author[Veropalumbo, A. et al.]{Alfonso Veropalumbo$^{1,2}$,\thanks{E-mail: alfonso.veropalumbo@uniroma3.it}
I\~nigo S\'aez Casares$^{3}$, 
Enzo Branchini$^{1,2,4}$,
\newauthor
Benjamin R. Granett$^{5, 6}$,
Luigi Guzzo$^{5, 6}$,
Federico Marulli$^{7, 8, 9}$,
Michele Moresco$^{7, 8}$,
\newauthor
Lauro Moscardini$^{7, 8, 9}$,
Andrea Pezzotta$^{10, 11, 12}$,
Sylvain de la Torre$^{13}$
\\~\\
$^{1}$Dipartimento di Matematica e Fisica, Universit\`a degli studi Roma Tre, Via della Vasca Navale, 84, 00146 Roma, Italy \\
$^{2}$INFN - Sezione di Roma Tre, via della Vasca Navale 84, I-00146 Roma, Italy\\
$^{3}$ \'Ecole Normale Sup\'erieure Paris-Saclay, 4 avenue des Sciences, 91190 Gif-sur-Yvette, France   \\
$^{4}$ INAF - Osservatorio Astronomico di Roma, via Frascati 33, I-00040 Monte Porzio Catone (RM), Italy\\
$^5$ INAF - Osservatorio Astronomico di Brera, Via Brera 28, 20122 Milano, via E. Bianchi 46, 23807 Merate, Italy\\
$^6$ Universit\`{a} degli Studi di Milano, via G. Celoria 16, 20133 Milano, Italy\\
$^{7}$ Dipartimento di Fisica e Astronomia ``Augusto Righi'' -
  Alma Mater Studiorum Universit\`{a} di Bologna, via Piero Gobetti
  93/2,\\ I-40129 Bologna, Italy\\
$^8$ INAF - Osservatorio di Astrofisica e Scienza dello Spazio
  di Bologna, via Piero Gobetti 93/3, I-40129 Bologna, Italy\\
$^9$ INFN - Sezione di Bologna, viale Berti Pichat 6/2,
  I-40127 Bologna, Italy\\
$^{10}$ Max-Planck-Institut f\"{u}r extraterrestrische Physik, Postfach 1312, Giessenbachstr., 85741 Garching, Germany\\
$^{11}$ Institute of Space Sciences (ICE, CSIC), Campus UAB, Carrer de Can Magrans, s/n, 08193 Barcelona, Spain\\
$^{12}$ Institut d`Estudis Espacials de Catalunya (IEEC), 08034 Barcelona, Spain\\
$^{13}$ Aix Marseille Univ, CNRS, CNES, LAM, Marseille, France\\
}

\date{Accepted XXX. Received YYY; in original form ZZZ}

\pubyear{2021}

\begin{document}
\label{firstpage}

\sloppy

\pagerange{\pageref{firstpage}--\pageref{lastpage}}
\maketitle

\begin{abstract}
We measure the galaxy 2- and 3-point correlation functions at $z=[0.5,0.7]$ and $z=[0.7, 0.9]$, from the final data release of the VIPERS survey (PDR2).  We model the two statistics including a nonlinear 1-loop model for the 2-point function and a tree-level model for the 3-point function, and perform a joint likelihood analysis. The entire process and nonlinear corrections are tested and validated through the use of the 153 highly realistic VIPERS mock catalogues, showing that they are robust down to scales as small as 10 $\Mpch$. The mocks are also adopted to compute the covariance matrix that we use for the joint 2- and 3-point analysis. Despite the limited statistics of the two (volume-limited) sub-samples analysed, we demonstrate that such a combination successfully breaks the degeneracy existing at 2-point level between clustering amplitude $\sigma_8$, linear bias $b_1$ and the linear growth rate of fluctuations $f$.
For the latter, in particular, we measure $f(z=0.61)=0.64^{+0.55}_{-0.37}$ and $f(z=0.8)=1.0\pm1.0$, while the amplitude of clustering is found to be $\sigma_8(z=0.61)=0.50\pm 0.12$ and $\sigma_8(z=0.8)=0.39^{+0.11}_{-0.13}$.
These values are in excellent agreement with the extrapolation of a Planck cosmology.  
\end{abstract}




\section{Introduction}
\label{sec:intro}

Galaxy clustering has emerged over the past three decades as one main observational pillar supporting the current standard model of cosmology. Cosmological constraints were so far derived mostly from two-point statistics of the galaxy distribution, either in configuration or in Fourier space \citep[see e.g.][for a comprehensive recent review]{Alam2021}. 
Two-point statistics will remain a central probe also for the next generation of redshift surveys, which have just started or are about to start. This includes in particular the measurement of specific features, as baryonic acoustic oscillations (BAO) \citep[e.g.,][]{Cole2005, Eisenstein2005}
and redshift space distortions (RSD) \citep[e.g.,][]{Peacock2001, Guzzo2008}, which are part of the standard ``dark energy" probes in projects like DESI \citep{Desi1, Desi2}, the Euclid \citep{EuclidRedbook} and Roman \citep{WFirst2019} space telescopes or, in the radio, the Square Kilometre Array \citep[SKA,][]{Ska}. 

For a Gaussian random field, this would be enough to fully characterize the field statistically.
However, significant information exists in the galaxy distribution, beyond the two-point
correlation function (2PCF hereafter) or the power spectrum. This is locked in the n-point correlation functions, or higher-moments, of the field. The simplest of these, the 3-point correlation function (3PCF hereafter), is  sensitive to non-Gaussian features in the primordial density perturbations,  non-linear effects in their evolution and 
galaxy biasing, i.e. the relationship between galaxy tracers and the underlying distribution of matter.
Since these effects are characterised by a different higher-order signal 
\citep[see e.g.][]{Fry1994,Matarrese1997}, the measurements of the 3PCF (or equivalently in Fourier space, the bispectrum) has the potential to separate them and constrain their relative strength and evolution. In addition, BAO and RSD also leave an imprint on higher-order statistics that can be exploited to remove parameter degeneracy and improve cosmological constraints.

The advantage of combining 2- and 3-point statistics to improve cosmological constraints has been realised in recent years \citep[][]{Sefusatti2006, Yankelevich2019}.
These works showed that the combination of the 3D power spectrum and bispectrum can significantly reduce statistical errors in the determination of cosmological parameters, while breaking degeneracies among these. The benefit is remarkable when using clustering data alone. These considerations have inspired the recent combined 2- and 3-point analyses on SDSS data carried out by \citet{GilMarin2015, GilMarin2017}.

In configuration space, the measurement of the 3PCF is computationally more demanding. Examples in the literature include estimates from the 2dFGRS \citep{Jing2004}, various releases of the SDSS main
sample \citep{Kayo2004,Nichol2006,Kulkarni2007,McBride2011a,McBride2011b,Marin2011,Guo2014}, the BOSS CMASS \citep{Guo2015} and the first Public Data Release (PDR1) of the VIPERS survey \citep{Moresco2017}.
Only recently, however, thanks to a breakthrough in the efficiency of available estimators, it has become possible to push measurements of the 3PCF to scales comparable to those of the BAO peak \citep{Slepian2015, Slepian2017c,Slepian2017a}, also using galaxy clusters as tracers
\citep{Moresco2020}.
Finally, combined 2PCF and 3PCF analyses in configuration space were performed by \citet{Marin2013} using the WiggleZ spectroscopic galaxy survey.

In this paper, we perform for the first time a joint 2- and 3-point correlation analysis of the final PDR2 catalogue of the VIPERS survey \citep{Scodeggio2018, Guzzo2014, Garilli2014}, applying the most advanced estimators and extracting joint cosmological constraints out to $z=0.9$.
One main  goal of VIPERS was to constrain the growth rate of structure $f$ out to $z\sim 1$, a result that was successfully achieved through a series of complementary estimates \citep{DeLaTorre2013, DeLaTorre2017, Pezzotta2017, Mohammad2018}.  All these analyses were developed in configuration space, for which the survey geometry and footprint can be more easily handled, if compared to the complex window function convolution one has to deal with in Fourier space \citep{Rota2017}. 
The main drawback when working in configuration space 
is the large covariance of 2PCF and 3PCF data and their errors. However, thanks to the availability of efficient clustering estimators and of the large ensemble of realistic mock catalogues made available by the VIPERS collaboration \citep{DeLaTorre2017}, we can solve the problem by numerically estimating the covariance matrix of the data.

We thus build upon previous VIPERS analyses to perform a joint 2PCF and 3PCF measurement, which allows us to break the degeneracy which affects some key cosmological parameters when these are estimated from the 2PCF alone. Specifically, we obtain  separate estimates of the linear growth rate of structure, $f$, the clustering amplitude, $\sigma_8$, and some of the parameters entering the galaxy biasing relation.
This work, in particular, expands upon the 3PCF analysis performed by \citet{Moresco2017} on the first data release of VIPERS (PDR1), by 1) using the final, larger PDR2 release; 2) exploring all triangle configurations, rather than just a subset;  3) combining 2- and 3-point statistics to obtain joint cosmological constraints.  It is also complementary to \citet{Cappi2015} and \cite{DiPorto2016}, where the non-linearity and evolution of galaxy bias in VIPERS was first studied.

The layout of the paper is as follows.
In Sect.~\ref{sec:data} we briefly describe the VIPERS PDR2 catalogue and 
the mock data used to estimate errors.
In Sect.~\ref{sec:estimators} we describe the 2PCF and 3PCF estimators and perform validation tests using the mock samples.
Covariance matrices and their estimates are discussed in Sect.~\ref{sec:covmatrix}.
In Sect.~\ref{sec:model} we present the 2PCF and 3PCF models  used for the likelihood analysis
of Sect.~\ref{sec:posterior}.
The results of the 2PCF, the 3PCF and their joint analyses are presented in 
Sect.~ \ref{sec:results}. Finally, we discuss the results and draw our conclusions in Sect.~\ref{sec:conclusions}.

Throughout the work, unless otherwise specified, we assume a flat $\Lambda$-cold dark matter ($\Lambda$CDM) cosmological model
characterized by the following parameters: $\lbrace \Omega_M, \Omega_b, n_s \rbrace = \left( 0.3, 0.045, 0.96\right)$.
The Hubble constant is defined as $H_0 = 100 \, h \, \mathrm{km}\, \mathrm{s}^{-1} \, \mathrm{Mpc}^{-1}$.

\section{Datasets}
\label{sec:data}

\subsection{VIPERS data}
\label{sec:VIPERS}

\begin{table*}\centering
\caption{Definition of the sub-samples of the VIPERS PDR2 catalogue that we have considered in this work.
    P1 and P2 are magnitude-limited samples that match those analysed in \citet{Pezzotta2017} and are used here for validation and consistency tests. G1 and G2 are instead volume-limited samples, upon which the joint clustering analysis specific of this work is based.}
{\renewcommand\arraystretch{1.2} 
\begin{tabular}{|c|c|c|c|}
\hline
Name & $z-$range & Magnitude Cut & Number of objects \\ \hline
G1 & $0.5 \leq z < 0.7$ & $M_B < -19.3+(0.7-z)$ & 23352 \\ \hline
G2 & $0.7 \leq z < 0.9$ & $M_B < -20.3+(0.9-z)$ & 13046 \\ \hline
P1 & $0.5 \leq z < 0.7$ & $i_{\rm AB} \leq 22.5$ & 30764 \\ \hline
P2 & $0.7 \leq z < 1.2$ & $i_{\rm AB} \leq 22.5$ & 35734 \\ \hline

\end{tabular}

\label{tab:samples}
}
\end{table*}

The VIMOS Public Extragalactic Redshift Survey (VIPERS) has been completed as one of the ESO Large Programmes. It was designed to build a spectroscopic sample of about 100,000 galaxies, aiming at an optimal combination of depth (reaching beyond $z\simeq 1$), volume and sampling density \citep{Guzzo2014}.  This is obtained covering $\sim$ 24 deg$^2$ over the W1 and W4 fields of the Canada-France-Hawaii Telescope Legacy Survey, which provides accurate photometry in five bands.
This area was tiled with a mosaic of 288 pointings with the VIMOS spectrograph at the ESO VLT,
measuring moderate-resolution spectra ($R\simeq220$) for galaxies brighter than $i_{\rm AB}=22.5$.  A 
colour pre-selection in the $(r - i)$ vs $(u - g)$ plane was applied prior to the spectroscopic observations, efficiently and accurately excluding objects with $z<0.5$ and boosting the spectroscopic sampling to nearly $50\%$. 
The root mean square ({\it rms}) redshift measurement error of these data is $\sigma_z=5 \times 10^{-4}(1+z)$, corresponding to 167 km s$^{-1}$. For consistency with previous clustering analyses of the VIPERS data, we only consider objects with a redshift confirmation rate 
larger than 96.1\%, corresponding to quality flags from 2 to 9. 
More details on the survey design and the final data release can be found in \citet{Guzzo2014} and \citet{Scodeggio2018}, respectively.

For the work presented here, we extract from the PDR2 catalogue four different sub-samples, whose characteristics are summarised in Table \ref{tab:samples}. The samples P1 and P2 are magnitude-limited samples corresponding to two redshift bins, $z=[0.5,0.7]$ and $[0.7, 1.2]$, following the selection criteria used in \citet{Pezzotta2017}. They will be used to validate our analysis against previous measurements from the same data. This selection maximises the number of available tracers, at the price of a redshift-dependent selection function and a mean density and bias that vary with redshift, especially in the outer redshift bin, which reaches $z=1.2$ with a rather sparse bright population.

The samples named G1 and G2 are those used for the specific new analyses of this paper. They are two volume-limited, non-overlapping bins at $z=[0.5,0.7]$ and $[0.7,0.9]$, coinciding with the samples named L1 and L3 in the SHAM analysis of PDR2 by \citet{Granett2019}, \citep[see also][]{Davidzon2016}. The absolute magnitude thresholds guarantee completeness above  90\%, while evolution (both in luminosity and bias) is minimised by the limited redshift size of the bins. 
Galaxies within G1 and G2 are brighter on average than those in P1 and P2.  As a consequence, the measured clustering amplitude
is expected to be larger due to the higher galaxy bias \citep{Marulli2013}, hence partly compensating for the larger Poisson noise. 
\subsection{Weights}
\label{sec:weights}
The completeness of the spectroscopic sample is quantified in the  PDR2 catalogue by a direction-dependent target sampling rate (TSR) and a spectroscopic success rate (SSR) \citep{Scodeggio2018}.  To account for these, we have used the same weighting scheme as in 
\citet{Pezzotta2017}. To each object, a statistical weight $w_i$ is assigned as
\begin{equation}
    w_i = w_i^{TSR} \cdot w_i^{SSR} \,\,\, .
    \label{eq:weights}
\end{equation}
Here, $w_i^{TSR}$ is defined  as  the  ratio  of  the  local  surface densities of target and parent galaxies (i.e.  before and after applying the target selection)
within an aperture of $60 \times 100$ arcsec$^2$.  $w_i^{SSR}$ is instead the local fraction of observed spectra with reliable redshift measurement with respect to the target sample. 
We also correct for the small-scale bias introduced by slit ``collisions" by up-weighting each galaxy-galaxy pair at a given angular separation according to \cite{DeLaTorre2013}. On the scales of our analysis, this last correction impacts the estimated 2PCF quadrupole moment, but not the monopole. We refer the interested reader to Sect.~4 of \citet{Pezzotta2017} for further details on the weighting scheme.

\subsection{Mock VIPERS data}
\label{sec:mocks}

To validate our analysis and quantify statistical errors,  we 
use the publicly available  VIPERS mock catalogues\footnote{The VIPERS mock catalogues are publicly available at \url{http://www.vipers.inaf.it/rel-pdr2.html\#mocks} }.
These were generated from lightcones extracted from
the Big MultiDark N-body simulation \citep{Klypin2016}. The cosmology of that simulations is a
flat $\Lambda$CDM characterised by the set of parameters $(\Omega_M, \Omega_\Lambda, \Omega_b, h, n_s, \sigma_8)=(0.307, 0.693, 0.0482, 0.678, 0.960, 0.823)$, which are
slightly different from those used in this work.
Dark matter halos were identified in the parent simulation and populated with synthetic galaxies down to the faint magnitude limit of the survey, as detailed in \citet{DeLaTorre2013, DeLaTorre2017}.
This resulted in 153 independent mock VIPERS W1+W4 catalogues, on which the same footprint and selection function of the real survey were applied. These mocks were designed to match the luminosity function, number density and redshift distribution of VIPERS galaxies. Previous analyses have shown that they also reproduce the VIPERS 2PCF
\citep{Granett2019}.  We will see in Sect.~\ref{sec:3PCF}  that, within the errors, they also match the 3PCF of VIPERS galaxies.

For each mock catalogue, four subsamples corresponding to the G1, G2, P1 and P2 selections of Table \ref{tab:samples} were created.

\section{Clustering Measurements}
\label{sec:estimators}

\begin{figure*}
	\includegraphics[width=\textwidth]{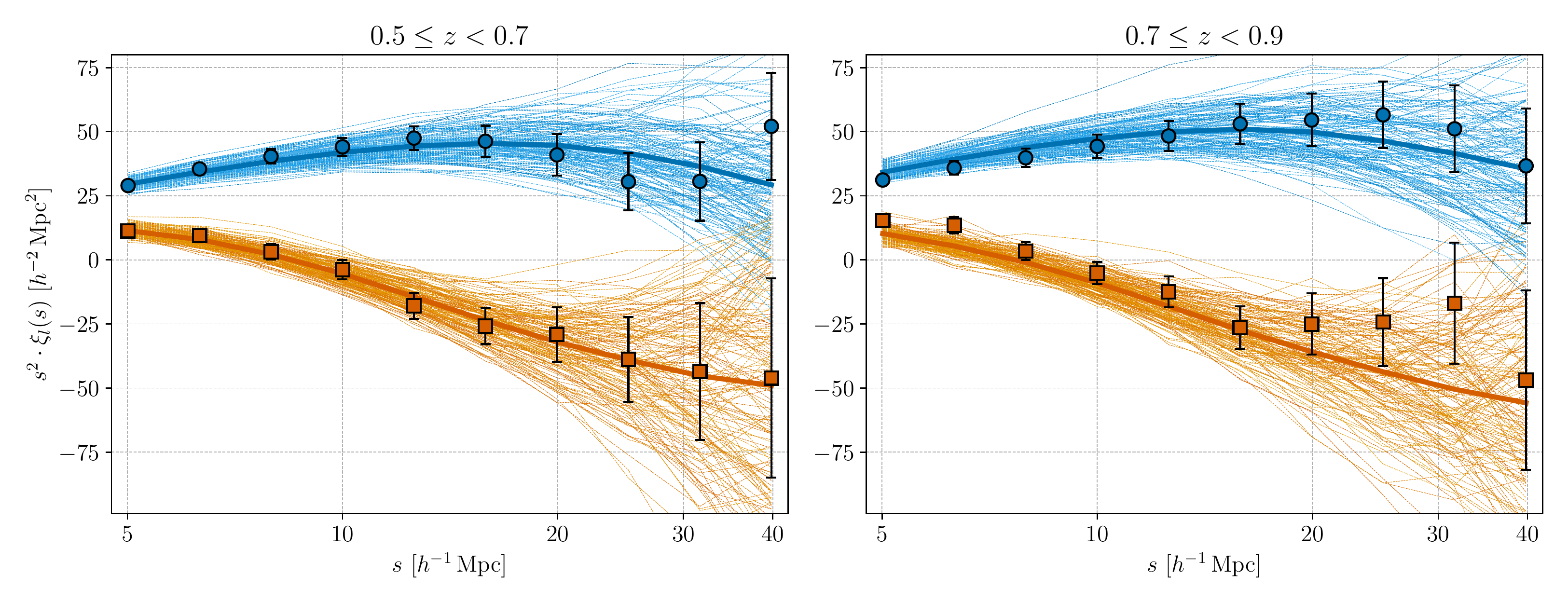}
    \caption{Monopole (blue circles) and quadrupole (orange squares) moments of the 2PCF measured from the G1 (left) and G2 (right) VIPERS sub-samples.  The corresponding measurements from the 153 VIPERS mock catalogues are also plotted (thin solid lines), together with their average (thick solid line). The error bars on the data points correspond to the {\it rms} dispersion of the mocks. The consistency between the mocks and the real data highlights the very good fidelity in clustering properties of the mock galaxies.}
    \label{fig:2pcf_multi}
\end{figure*}

The estimators used in this work belong to the class of unbiased, minimum variance N-point estimators proposed by \citet{Szapudi1998}. Their general form is

\begin{equation}
    \xi^N = \frac{ (D - R)^N }{R^N},
    \label{eq:xiN}
\end{equation}
where $D$ is the data catalogue, $R$ is the so-called ``random" sample, and $N$ is the order of the correlation statistics. The estimate relies on counting and binning all N-tuples $D^q R^p$ formed by $q$ data and $p$ random objects, with $q+p = N$. 

The "random" catalogue of objects is a synthetic sample with the same geometry and selection function as the real survey but with no spatial clustering.
We assume that the selection function, i.e. the probability
to observe a galaxy at the spatial position  $(\alpha, \delta, z)$, can be factorised as
\begin{equation}
    P(\alpha, \delta, z) = f(\alpha, \delta) N(z) \, , \label{eq:sel_func}    
\end{equation}
where $f(\alpha, \delta)$ accounts for the angular footprint of the survey and $N(z)$ is the redshift
distribution of the sources in the catalogue.

All measurements presented here have been performed using the estimators described below and publicly available in the library
{\small CosmoBolognaLib}\footnote{\url{https://gitlab.com/federicomarulli/CosmoBolognaLib}} \citep{Marulli2016}.

\subsection{2PCF estimator}
\label{sec:2PCF}

\begin{figure*}
	\includegraphics[width=\textwidth]{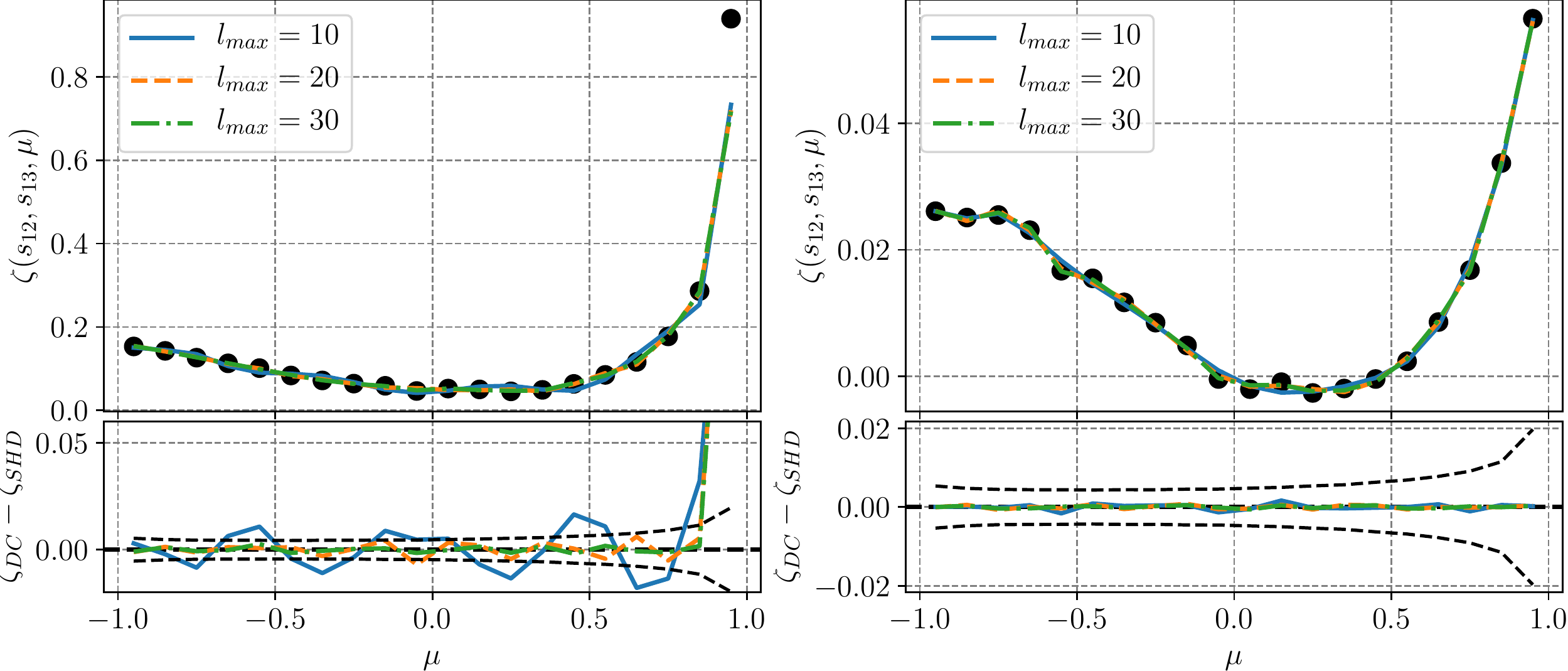}
    \caption{Top panels: estimates of the three-point correlation function for the G1 sample using both the triplet counting method (black dots) and  the SHD method (solid lines), for different $l_{max}$ as indicated by the legend (top panels). The left and right panels correspond to isosceles  ($r_{12} = r_{13} = 10  \, \Mpch$) and non-isosceles 
    ($r_{12} = 10 \, \Mpch, r_{13} = 30 \, \Mpch$) configurations, respectively.  The bottom panels show the difference between the two estimators, for the different choices of $l_{max}$ (solid lines). The dashed black curves  correspond to $\pm 10\%$ of the statistical scatter of the mocks, highlighting the relative importance of the systematic errors introduced by the choice of $l_{max}$. 
}
    \label{fig:dir_vs_shd}
\end{figure*}

To estimate the anisotropic 2PCF we use the  \citet{Landy1993} estimator;
\begin{equation}
    \xi (s, \mu) \, = \frac{DD(s, \mu)-2DR(s, \mu)-RR(s, \mu)}{RR(s, \mu)},
    \label{eq:xi_LS}
\end{equation}
where $DD$, $DR$ and $RR$ are the data-data, data-random and random-random pairs of objects, respectively. The separation vector of two objects has modulus $s$
and forms a  cosine angle $\mu = \cos(\theta)$ with the line of sight to the pair, measured at the midpoint of the separation vector itself.

We bin the pair counts using a constant logarithmic bin in $s$, $\Delta \log(s) = 0.1$,
and a linear one for $\mu$, $\Delta \mu = 0.05$, as in 
\citet{Pezzotta2017}.

For each $s$ bin we compute the 2PCF multipoles
\begin{equation}
    \xi_l (s) = \frac{2l+1}{2}\int_{-1}^1 \de \mu \Leg{l}{\mu} 
                \xi(s, \mu) \, ,
    \label{eq:xil}
\end{equation}
where $ \Leg{\ell}{\mu}$ are the Legendre polynomials. 
We only consider the monopole $\ell=0$ and quadrupole $\ell=2$, since
odd multipoles are zero by design, and the measured hexadecapole turns out to be consistent with zero within the (large) errors.

To validate our estimator, we measure the 2PCF multipoles for the subsamples of 
 Table \ref{tab:samples} and compare the results with those of previous VIPERS studies. 
Fig.~\ref{fig:2pcf_multi} shows the 2PCF monopole and quadrupole moments of the VIPERS G1 and G2 samples (blue circles and the orange squares, respectively).
The solid thin lines correspond to measurements of each mock catalogue, with the thicker line and errorbars on data points showing the average and {\it rms} of the mocks respectively.
This plot shows how the mock VIPERS samples reproduce faithfully the clustering of the real data, suggesting that they are adequate to estimate their errors and covariance. We have repeated the test using the  P1 and P2 samples and verified that the results match those of \citet[][see Fig.~12]{Pezzotta2017}

The maximum separation considered in our analysis is set by the signal-to-noise of the 2PCF.
The VIPERS footprint is highly elongated along the (equatorial) longitude, which significantly reduces the number of distant pairs in the transverse direction. As a result, the estimated quadrupole moment becomes noisy
beyond $\sim 40 \, \Mpch$. We therefore set  $s_{max}=40 \, \Mpch$ in our 2PCF analysis.
In Sect.~\ref{sec:2pt_results} we will also test the sensitivity of the results to the smallest separation scale considered, showing that the results are robust down to $s_{min}= 15 \, \Mpch$, which we then assume as the default value. In summary, our baseline range for the 2PCF analysis is $[15,40] \, \Mpch$. 

\subsection{3PCF estimator}
\label{sec:3PCF}

\begin{figure*}
	\includegraphics[width=\textwidth]{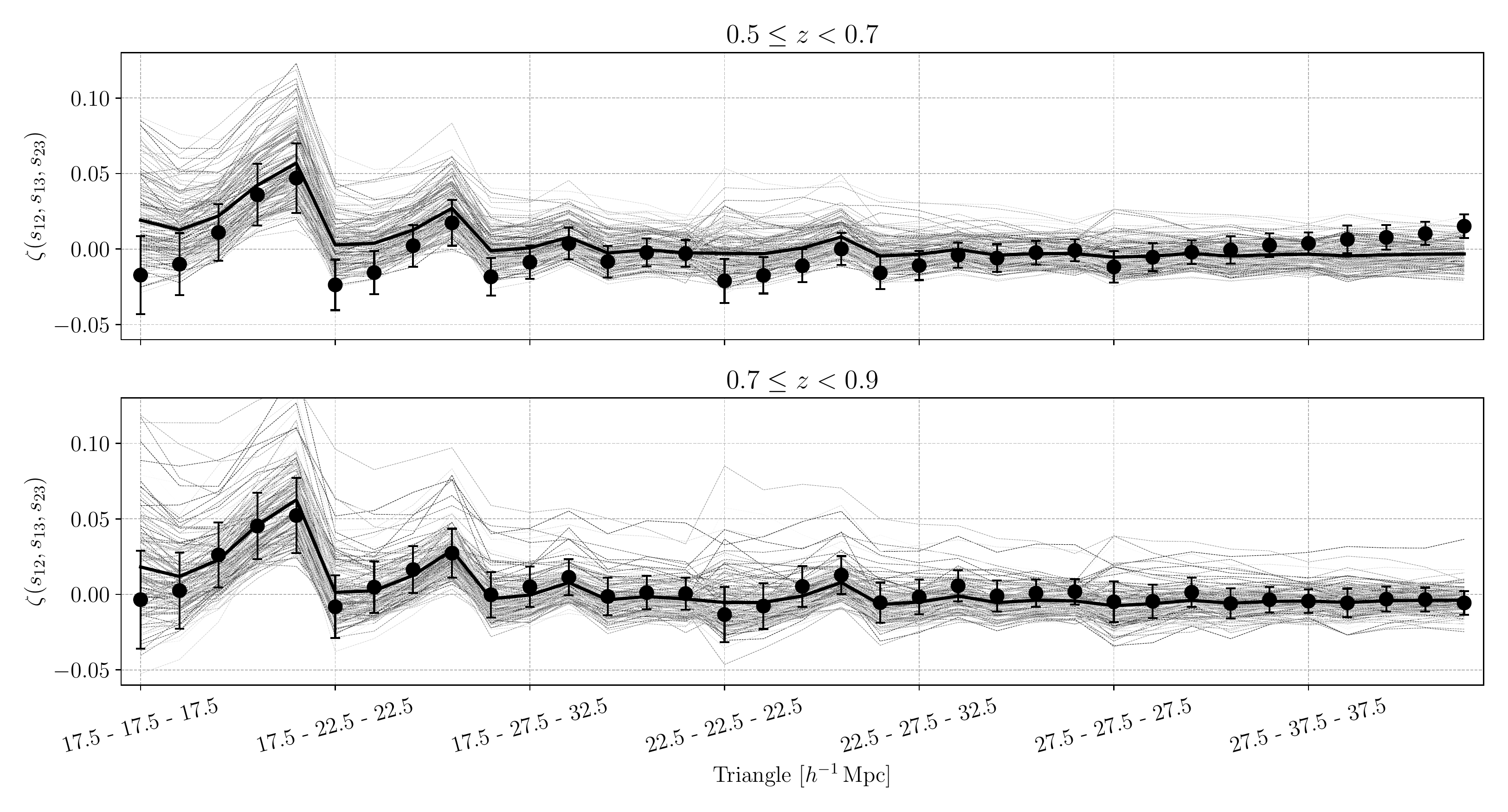}
    \caption{Monopole moments (black circles) of the three-point correlation function for the G1 and G2 samples (top and bottom panels, respectively), measured  for  some  selected  sets  of  triangles  of different side lengths, as indicated on the x-axis. The  corresponding  measurements  from  the  153  VIPERS  mock  catalogues  are  also  shown by the thin grey lines, together with their averages (solid black line).  Error bars correspond to the {\it rms} scatter among the mocks.
    }
    \label{fig:3pcf_mono}
\end{figure*}

To measure the 3PCF of the VIPERS first data release catalogue (PDR1), \citet{Moresco2017} used the estimator of \citet{Szapudi1998} because of its ability to account for complicated survey geometry  \citep{Kayo2004}. 
This estimator, however, is computationally demanding since it relies on brute-force galaxy triplet counting, whose computing time scales as $O(N^3)$. 
This becomes unbearable if applied to 153 mock catalogues with the size of VIPERS.

A major breakthrough in this respect has been the introduction by \citet{Slepian2015}, of a more efficient estimator based on a spherical harmonics decomposition (hereafter SHD).
For this algorithm, the computational cost scales as $O(N^2)$, making it successfully applicable to large galaxy samples such as the SDSS-DR12 CMASS catalogue \citep{Slepian2017a}. The method relies on the 3PCF Legendre polynomials expansion proposed by \citet{Szapudi2004}
\begin{equation}
    \zeta (s_{12}, s_{13}, \mu) \, = \sum_{l=0} ^ {l_{max}} \zeta_{l}(s_{12}, s_{13}) \Leg{l}{\mu}\, ,
   \label{eq:zeta_from_leg}
\end{equation}
where the 3PCF $\zeta(s_{12}, s_{13}, \mu)$ is parametrized by two triangle sides $s_{12}$, $s_{13}$ and their cosine angle $\mu=\cos(\hat{s}_{12} \cdot \hat{s}_{13})$.
This expansion offers two advantages. The first one is that the multipole moments $\zeta_{l}(s_{12}, s_{13})$ can be efficiently estimated by locally expanding the density field at distances $s_{12} $ and $ s_{13}$ in spherical harmonics and then by cross-correlating the expansion coefficients using the spherical harmonics addition theorem. 
The second advantage is that Eq. \eqref{eq:zeta_from_leg} typically requires a limited number of multipoles to converge.
These properties dramatically reduce the computational cost, allowing us to measure the 3PCF in all VIPERS mocks and consider all triangle configurations.
We refer the reader to \citet{Slepian2015} for a detailed description of the SHD estimator.

The main disadvantages of this estimator are that, for a given configuration, all triangles are mixed together;
this is particularly relevant in all cases where the third side spans a large range of scales. For example, in the case of isosceles configurations with $s_{12} = s_{13}$, the third side $s_{23}$
varies from 0 to $2 \times s_{12}$, i.e. well into the highly nonlinear regime,
where theoretical predictions cannot be trusted. For this reason, here we will only consider triangles with all side lengths above a minimum
value $s_{min}=15 \, \Mpch$ and up to a maximum length $s_{max}=40 \, \Mpch$.
This matches the scale used in the 2PCF analysis.
In Sect.~\ref{sec:3pt_results}, we shall test the robustness of our results to such a choice.

\subsubsection{Convergence of the SHD method}
\label{sec:3pcf_convergece}

We first assess the sensitivity of the SHD method to the choice of $l_{max}$ in Eq.~\ref{eq:zeta_from_leg}.
We thus compare the 3PCF of the G1 catalogue
measured with the SHD method to that, supposedly exact, estimated with the same brute-force triplet counting technique used by \citet{Moresco2017}. 
We show in Fig.~\ref{fig:dir_vs_shd} the results for two triangle configurations: isosceles triangles with $s_{12} = s_{13} = 10 \, \Mpch$ (left panel)
and non-isosceles triangles with $s_{12} = 10 \, \Mpch$ and $ s_{13} = 30  \, \Mpch$ (right panel). In both cases, the same $\mu$ binning is used.

In the top panels, we compare the reference 3PCF measured with the triplet counting technique (black dots), with the one measured with the SHD method using
$l_{max} = 10, \, 20$ and $30$ (lines with different colours). These particular choice of configurations cover both the cases where highly non-linear scales are included (isosceles, left panel) or not (non-isosceles, right panel).
In the bottom panels, we show the difference of the 3PCFs estimated with the two methods. As an indication of the relative value of the potential systematic error with respect to statistical errors, the black dashed lines indicate 
10\% of the {\it rms} scatter among the mock catalogues. For non-isosceles configurations (right panel), the SHD matches well the expected result, with differences of the order (or below) 1\% of the random errors, almost independently of the choice of $l_{max}$. This implies that a limited number of multipoles $\ell$ is sufficient for  Eq. \eqref{eq:zeta_from_leg} to converge to the correct result, typically $l<10$.

On the contrary, the results of the SHD method for the isosceles configurations (left panel)
depend on the choice of $l_{max}$ and $\mu$. 
For $l_{max}=30$, the results are generally satisfactory (the difference between SHD and triplet counting is of the order of a few percent of the statistical error), except for $\mu=1$, i.e. when $s_{23} \rightarrow 0$.
This is not surprising, since elongated isosceles triangle shapes are difficult to reproduce in harmonics space.

Based on these results, we adopt a conservative approach in which: 1) we set $l_{max} = 30$ in the 3PCF estimate, and  2) we exclude the case $\mu=1$ for isosceles configurations. 
We also choose to parameterise the 3PCF
in terms of the three side lengths $\zeta (s_{12}, s_{13}, s_{23})$ instead of two sides and the cosine angle, $\zeta (s_{12}, s_{13}, \mu)$.
Since the SHD method involves binning in the radial direction,
we perform an additional step, described in Appendix \ref{app:3pcfest}, to 
use the same binning along the three sides $s_{12}, s_{13}$ and $s_{23}$.

Finally, we sort the triangles in increasing size $s_{12} \leq s_{13} \leq s_{23}$, to avoid repetitions.
This choice also facilitates triangle selection:
for example, we can exclude scales smaller than a chosen value by setting a threshold for one triangle side only. 
 
\subsubsection{Comparison with the 3PCF of mock samples}
\label{sec:3pcf_mock_comparison}

The second test is analogous to the one performed to validate the 2PCF estimator: we measure the 3PCF in the G1, G2, P1 and P2 VIPERS samples and compare the results with the same quantity measured in the mocks. 
 
In Fig.~\ref{fig:3pcf_mono} we show the results for the G1 and G2 samples (top and bottom panels, respectively). 
We detect a non-zero 3PCF signal up to the largest scales considered in our analysis, in agreement with  \citet{Moresco2017}. The VIPERS 3PCF agrees well with that of the mock samples, demonstrating that the VIPERS mocks are indeed adequate to estimate the errors for the 3PCF analysis, too.

\section{Covariant errors}
\label{sec:covmatrix}

\begin{figure}

	\includegraphics[width=0.5\textwidth]{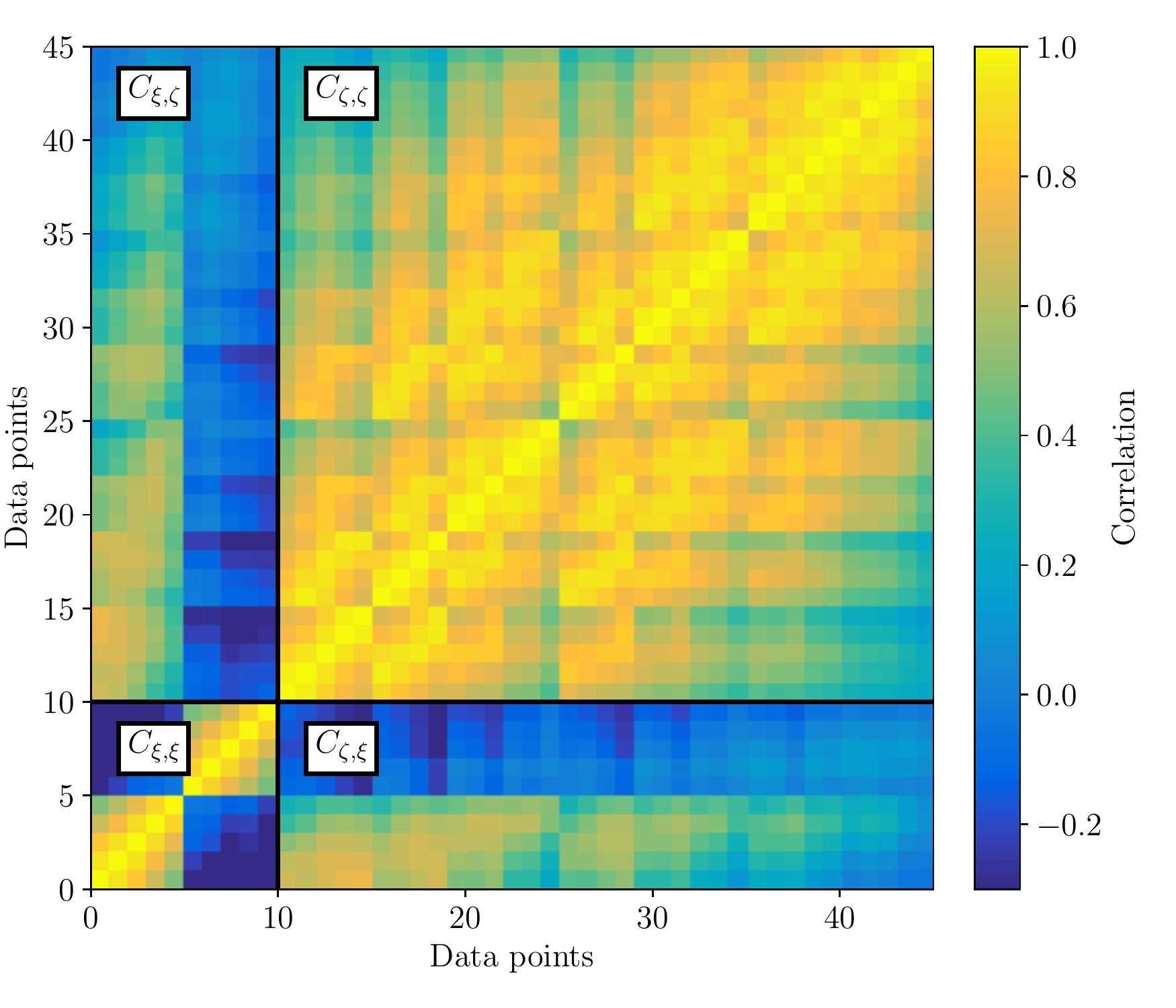}
	\caption{The global correlation matrix used in our joint likelihood analysis of the 2PCF and 3PCF, estimated from the 153 VIPERS mock catalogues using Eq.~\ref{eq:bf_cov}. We show here the one for the G1 sample, as an illustrative example. The matrix is composed by two sub-blocks, corresponding to the 2PCF monopole and quadrupole ($C_{\xi, \xi}$, bottom left), and the 3PCF monopole ($C_{\zeta, \zeta}$, top right), respectively, together with their cross-covariance ($C_{\xi, \zeta}$).  The sub-arrays are used for the separate 2PCF and 3PCF analyses, while the full $45\times45$ matrix is used in the joint likelihood analysis.
	}
    \label{fig:joint_corr}
\end{figure}

To estimate the errors, we will not adopt an analytic, Gaussian model (like in, e.g., \citealt{Slepian2017b}). Instead, supported by the results of the previous section, we directly estimate errors and their covariance matrix $C_{i, j}$
from the mock VIPERS catalogues, as

\begin{equation}
	\label{eq:bf_cov}	
    C_{i, j} = \frac {1}{N_{m}-1} \sum_{k=0}^{N_{m}} \left( d_i^k-\overline{d}_i \right)
               \left( d_j^k-\overline{d}_j  \right) \, ,
\end{equation}
where $d_i^k$ indicates the data vector in the $i-{th}$ bin and in the $k-{th}$ mock catalog. The size of the data vector and of the covariance matrix depend on the specific clustering analysis being performed.
In the 2PCF analysis, performed in the range $15 \leq s_{min}^{2pt}  \leq 40 \, \Mpch$,
the size of the data vector is 10 and the corresponding $10\times10$ covariance matrix is shown in the lower left part of Fig.~\ref{fig:joint_corr}. For the 3PCF analysis, performed over the same range of scales, the size of the data vector is 35 and the covariance matrix corresponds to the upper-right part of the same figure. 
In the joint 2PCF and 3PCF analysis, the data vector has 45 elements and the covariance matrix, which also contains the  2PCF-3PCF cross-terms, corresponds to the full matrix of Fig.~\ref{fig:joint_corr}. 

Both precision and accuracy of our covariance matrix
depend on the number of mock catalogues used for its numerical estimation.
Since this number is limited, a correction for the expected systematic errors should be considered \citep[see, e.g. discussion in][]{Hartlap2007, Percival2014}. To this end, we follow \citet{Sellentin2016} in using a modified likelihood function, which we discuss in detail in Sect.~\ref{sec:posterior}.

\section{Modelling the galaxy 2- and 3- point correlation functions}
\label{sec:model}

In this section we describe our model for the anisotropic galaxy 2PCF, 
\begin{eqnarray}
    \label{eq:cl_2pcf_def}
    \xi_g(s, \mu) \, & = & \left< \delta_g(\vec{x})) \delta_g(\vec{x}+\vec{s}) \right> ,
\end{eqnarray}
and 3PCF,
\begin{eqnarray}
    \label{eq:cl_3pcf_def}
     \zeta_g(s_{12}, s_{13}, s_{23}) & = &\left< \delta_g(\vec{x}) \delta_g(\vec{x}+\vec{s_{12}}) \delta_g(\vec{x}+\vec{s_{13}}) \right> \, .
\end{eqnarray}
In these expressions, $\delta_g(\vec{x})$ is the galaxy density
contrast at the position $\vec{x}$, while
 $<...>$ indicates ensemble average.

The model relies on the approach of 
\citet{Scoccimarro1999, Scoccimarro2004, McDonald2009, Saito2014}, and has already been  
used to perform 2PCF, 3PCF and joint 2PCF+3PCF 
analyses \citep{GilMarin2015, DeLaTorre2017, Slepian2017b}.

\subsection{Galaxy bias}

The first ingredient of the model is 
the galaxy biasing relation of \citet{McDonald2009}  

\begin{eqnarray}
    \label{eq:delta_g}
     \delta_g  = b_1 \delta + 
    \frac{b_2}{2} \left[ \delta^2 - <\delta^2> \right] + 
  \frac{b_{s^2}}{2} \left[ s^2 - <s^2> \right] +  O\left(\delta^3\right).  \, \, 
\end{eqnarray}
In this expression, the galaxy density contrast $\delta_g$ is a function of the matter non-linear over-density $\delta$ and of the non-local tidal field $s^2$, which accounts for dependence of $\delta_g$ on the local potential $\phi$ \citep[see e.g. Sect.~II of ][]{McDonald2009}. The terms $<\delta^2>$ and $<s^2>$ guarantee that $<\delta_g>=0$. This bias model is then characterised by the four parameters $b_1$, $b_2$, $b_{s^2}, b_{3nl}$.

\subsection{Distortion effects}

Using the observed spectroscopic redshifts as a distance proxy introduces spurious distortions in the clustering properties that need to be accounted for.  Three types of distortions need to be considered: those introduced by the departures from the Hubble flow, those induced by assuming an incorrect cosmological model in computing distances from redshifts and those due to the errors on the measured redshifts.

\subsubsection{Peculiar motions}

Peculiar velocities introduce a Doppler shift that modifies the mapping between spatial positions, $\vec{x}$, and the measured positions, $\vec{s}$:
\begin{equation}
    \label{eq:rsd_mapping}
    \vec{s} = \vec{x} + \frac{v_{\parallel}}{a H(a)} \vec{e}_{\parallel} \,
\end{equation}
where $a$ is the expansion factor, $v_{\parallel}$ is the radial component of the peculiar velocity vector, $\vec{e}_{\parallel}$ is the unit radial vector,
and $\vec{s}$ is the redshift-space vector position of the object.

This mapping introduces a relation between 
the measured (redshift-space) galaxy over-density, $\delta_g (\vec{s})$, and the true, real-space one, $\delta_g(\vec{x})$
\begin{equation}
    \label{eq:rsd_density}
    \delta_g(\vec{s}) = \left[1+ \delta_g(\vec{x}) \right]
    \left|\frac{\mathrm{d}^3 s}{\mathrm{d}^3 x}\right|^{-1} -1\, ,
\end{equation}
where $\left|\frac{\mathrm{d}^3 s}{\mathrm{d}^3 x}\right|^{-1}$ is the Jacobian of 
the map in Eq.~\eqref{eq:rsd_mapping}.

In this work, we make the plane-parallel hypothesis, i.e. we assume that the relative separations between galaxy pairs are much smaller than the distance to the observer. This assumption is fully justified, since 
 the maximum scale considered in our analyses ($40 \, \Mpch$) is much smaller than any distance to a VIPERS galaxy, which lies at $z\geq0.5$.  Under this approximation, Eq.~\eqref{eq:rsd_density} becomes
\begin{equation}
    \label{eq:rsd_density_ppa}
    \delta_g(\vec{s}) = \frac{\delta_g(\vec{x})+f\partial_{\parallel}\vec{u}}{1-f\partial_{\parallel}\vec{u}} \, ,
\end{equation}
where the partial derivative is taken along the radial direction, $\vec{u}$ is the peculiar velocity, and $f$ is the linear growth rate of density fluctuations,
\begin{equation}
    \label{eq:linear_growth_rate}
    f = \frac{\mathrm{d} \log D}{\mathrm{d} \log a} \approx \Omega_M^{0.545}(z) \, ,
\end{equation}
where $D(z)$ is the linear growth factor and the second relation is a good approximation in the flat $\Lambda$CDM model assumed here \citep[see e.g.][]{Wang1998, Huterer2007}.

\subsubsection{Geometrical distortions}

Choosing an incorrect fiducial cosmology to estimate distances from redshifts generates 
another kind of anisotropy, the well-known Alcock-Paczynski effect \citep{AP79}.
The detection of this effect on intrinsically isotropic clustering features as the BAO peak in the 2PCF has been used to trace the expansion history of the Universe \citep{Eisenstein2005, Kazin2013}. 
Here, we will treat this effect as a nuisance to be marginalised over, since our analysis is limited to scales well below the BAO peak.  As shown in \citet{DeLaTorre2017}, the impact of this marginalisation in the error budget of the VIPERS clustering analysis is negligible. Therefore,
we will set our fiducial cosmology equal to the flat $\Lambda$CDM model specified in Sect.~\ref{sec:intro}
and safely ignore the  Alcock-Paczynski effect.

\subsubsection{Redshift measurement errors}

Redshift measurement errors introduce a noise in the redshift-to-distance relation analogous to that of incoherent motions \citep{Marulli2012}, effectively erasing information 
 on scales below
\begin{equation}
    \label{eq:zerr}
    \sigma_{\pi} = \frac{c \sigma_z}{H(z)} \, \Mpch\, ,
\end{equation}
where $\sigma_z$ is the $rms$ error typical of the adopted instrumental setup. For VIPERS, the errors are very well described by a Gaussian distribution with $\sigma_z = 5 \times 10 ^ {-4} (1+z)$, corresponding to a length scale $\sigma_{\pi} \sim 1.5 \, \Mpch $ \citep{Scodeggio2018, Sereno2015}.

\subsection{2PCF model}
\label{sec:2pointm}

To model the galaxy 2PCF we start from the matter power spectrum
$P(k)$
\begin{equation}
    \label{eq:pk_def}
    \left< \delta(\vec{k}) \delta(\vec{k^\prime}) \right> = 2 \pi^3 \delta_D(\vec{k}-\vec{k^\prime}) P(|\vec{k}|) \, .
\end{equation}
Using  Eq.~\eqref{eq:rsd_density_ppa} to account for redshift space distortions, we obtain an expression of the redshift-space power spectrum
\begin{eqnarray}
    \label{eq:pk_redshift_space}
    P^s(k, \nu) & = & \int d^3 \vec{r} 
    e^{-i\vec{k}\cdot\vec{r}} \left< e^{-i f k \nu \Delta u_{\parallel} } \right. \\ \nonumber
    & & \left . \left[ \delta(\vec{x}) + f\partial_{\parallel} u_{\parallel}    (\vec{x}) \right] \left[ \delta(\vec{x}^{\prime}) + f\partial_{\parallel} u_{\parallel} (\vec{x}^{\prime}) \right] \right>\, ,
\end{eqnarray}
where $\nu = k_{\parallel}/k$, $u_{\parallel} = - v_{\parallel}/(f a H(a))$, $v_{\parallel}$ and $k_{\parallel}$ are the
radial components of the peculiar velocity and wavenumber vectors, respectively, $\delta$ is the matter over-density, $u_{\parallel} = u_{\parallel}(\vec{x}) - u_{\parallel}(\vec{x}^{\prime})$ and
$\vec{r} = \vec{x}-\vec{x}^{\prime}$.
The term in square brackets accounts for the effect of coherent motions that increase the clustering amplitude, whereas the exponential factor encodes the ``Fingers of God" effect of incoherent motions. For a more detailed description, see \citet{Taruya2010}.

Equation \eqref{eq:pk_redshift_space} is exact but of impractical use. Rather, using the approximations introduced by \citet{Scoccimarro2004} and \citet{Taruya2010} for the case of biased mass tracers, we obtain the simpler expression

\begin{equation}
    P_g^s(k,\nu) = D(\sigma_{12}, \sigma_z)\left[\Pgg(k)+2\nu^2 f \Pgt(k) + \nu^4 f^2 \Ptt(k) \right] ,
    \label{eq:pgg}
\end{equation}
where
\begin{eqnarray}
  P_{gg}(k) = & b_1^2 P_{\delta\delta}(k)+2b_2b_1P_{b2,\delta}(k)+2b_{s^2}b_1P_{bs2,\delta}(k) \nonumber \\
  & + b_2^2P_{b22}(k) +2b_2b_{s^2}P_{b2s2}(k)+b_{s^2}^2P_{bs22}(k) \nonumber \\
  & + 2b_1b_{3\rm nl}\sigma_3^2(k)P_{\rm lin}(k), \\
  P_{g\theta}(k) =&b_1P_{\delta\theta}(k)+b_2P_{b2,\theta}(k)+b_{s^2}P_{bs2,\theta}(k) \nonumber \\
  & +b_{3\rm nl}\sigma_3^2(k)P_{\rm lin}(k) \, .
\end{eqnarray}
In these equations, 
$\Pdd$, $\Pdt$ and $\Ptt$
are, respectively, the nonlinear matter density-density,
density-(velocity divergence), and (velocity divergence)-(velocity
divergence) power spectra. $P_{\rm lin}$ is the
matter linear power spectrum, and $P_{b2,\delta}$, $P_{bs2,\delta}$,
$P_{b2,\theta}$, $P_{bs2,\theta}$, $P_{b22}$, $P_{b2s2}$, $P_{bs22}$,
$\sigma_3^2$ are 1-loop bias integrals; their expressions can be found in e.g. \citet{GilMarin2015, DeLaTorre2017}.
The linear and nonlinear power spectra of density and velocity divergence are
specified at the effective redshift of the sample. 
To model $P_{\rm lin}$ 
 we use the {\small CAMB} Boltzmann solver code \citep{Lewis2000},
whereas for  $P_{\delta 
\delta}(k), P_{\delta \theta}, P_{\theta \theta}$
we use the SPT 1-loop model of \citet[][Eqs. 63-65]{Scoccimarro2004}. 
In this model, the degree 
of non-linearity in $\Pdd$, $\Pdt$ and $\Ptt$ is quantified by the same parameter that measures the clustering amplitude, i.e. $\sigma_8(z)$.

In Eq. \eqref{eq:pgg} we have introduced the term $D(\sigma_{12}, \sigma_z)$ to model the combined damping effect of incoherent motions and redshift errors. Its explicit expression is
\begin{equation}
  \label{eq:nonlinear_rsd_damping}
  D(\sigma_{12}, \sigma_z)= \left(1+k^2\nu^2\sigma^2_{12}\right)^{-1} \cdot \exp \left(- k^2\nu^2 \sigma^2_{z}\right) \, ,
\end{equation}
The first term, modelled as a Lorentzian damping, accounts for the effect of random motions within dark matter halos, quantified by the pairwise velocity dispersion $\sigma_{12}$. The second term accounts for the redshift measurement errors and is modelled as a Gaussian function.
In our analysis, we fix $\sigma_z$ to the VIPERS estimated value and leave $\sigma_{12}$ as a free factor we can fit for and then treat it as a nuisance parameter.

Since the 2PCF model fits well the corresponding measurements from the VIPERS mocks 
on all scales considered, we decided to ignore the model correction terms $C_A(k,\nu,f,b_1)$ and $C_B(k,\nu,f,b_1)$ of \citet{Taruya2010}, that are instead considered in other analyses \citep[e.g.][]{GilMarin2015}.

To test the robustness of our model, we also considered an alternative 2PCF model, in which we used (1) the {\sc HALOFIT}  semi-analytical prescription
calibrated by \citet{Takahashi2012} to model the matter power spectra and, (2) the fitting functions of \citet{Bel2019}
to model $P_{\delta, \theta}(k)$ and $P_{\theta,\theta}(k)$. We find that the alternative model provide results very similar to the original one, the differences being much smaller than the statistical errors.

We then use Eq. \eqref{eq:pgg}
to model the 
multipole moments of the anisotropic power spectrum
\begin{equation} \label{expmomK}
P^s_\ell(k)=\frac{2\ell+1}{2} \int_{-1}^1 P_g^s(k,\nu) \Leg{l}{\nu} \,\d\nu
\end{equation}
and, from these, the multipole moments of the anisotropic 2PCF
\begin{equation} \label{expmom}
\xi^s_\ell(s)=i^\ell \int \frac{k^2}{2\pi^2} P^s_\ell(k)j_\ell(ks)\,\d k ,
\end{equation} 
where $j_\ell$ indicates the spherical Bessel functions.
Having neglected the  $C_A(k,\nu,f,b_1)$ terms, in our model
the parameters $b_1$ and $f$ are fully degenerate with $\sigma_8$. 
The two bias parameters $b_1$ and $b_2$ are also highly degenerate.

\subsection{3PCF model}
\label{sec:3pointm}
 The 3PCF model we adopt is the same one used by \citet{Slepian2017b} and \citet{Slepian2017c}
to detect the BAO feature in the 3PCF of SDSS DR12 galaxies and to measure their biasing relation in redshift space.

Our 3PCF model is
derived from galaxy bispectrum $B(|\vec{k_1}|, |\vec{k_2}|, |\vec{k_3}|)$, defined as

\begin{equation}
    \label{eq:bisp_fund}
     \left< \delta_g(\vec{k_1}) \delta_g(\vec{k_2}) \delta_g(\vec{k_3}) \right> = \delta_D(\vec{k_1}+\vec{k_2}+\vec{k_3}) B(|\vec{k_1}|, |\vec{k_2}|, |\vec{k_3}|).
\end{equation}
In particular, we assume the redshift-space tree-level galaxy bispectrum derived in \citet{Scoccimarro1999}.
From this, we obtain the 3PCF model by Fourier transform, following the prescriptions of \citet{Slepian2017a}. 

The 3PCF model can be expressed as
\begin{equation}
\begin{aligned}
    \zeta (s_{12}, s_{13}, s_{23})  = &
    \sum_{l=0}^{l=4} A_l (b_1, \gamma, \gamma^{\prime},
    \beta) \cdot
    f_l \left(s_{12}, s_{13}, s_{23}\right) + \\
    &  B(b_1, \beta) \cdot \sum_{l=0}^{l_{max}} 
    k_l \left( s_{12}, s_{13}, s_{23} \right) \, ,
\end{aligned}
\label{eq:zeta_model_compact}
\end{equation}
where $\gamma\equiv {b_2}/{b_1}$ and $\gamma^{\prime}\equiv b_{s^2}/b_1$ are combinations of the bias parameters.
The first term of Eq. \eqref{eq:zeta_model_compact} accounts for contributions separable in $k_1, k_2$ in
the \citet{Scoccimarro1999} bispectrum model. Its Fourier transform
can be computed analytically as the product of two quantities.
The first one, $A_l$, depends on the bias parameters, $b_1, b_2$, $b_{s^2}$, and on the linear distortion parameter $\beta=f/b_1$. The second one, 
 $f_l$, depends on the linear power spectrum, $P_{\rm lin}(k)$. The explicit expression of the 
$A_l$ and $f_l$ terms is given in  Appendix \ref{app:AlBl}.

The second term in Eq. \eqref{eq:zeta_model_compact} accounts for the non separable terms (i.e. the terms in which $k_3$ appears explicitly). It is also made-up of two terms. The multiplicative factor $B = b_1 ^3 ( 7 \beta^2 + 2 \beta^3)$
depends on $b_1$ and $\beta$ while the terms  $k_l \left( s_{12}, s_{13}, s_{23} \right)$ in the sum depend on $P_{\rm lin}(k)$.
Their explicit expressions are given in Equations 17, 18 and 19
of \citet{Slepian2017a}. 
The contribution of non-separable terms is small compared to that of the separable one and therefore we will ignore them in the 3PCF model, similarly to what we did with the $C_A$ and $C_B$ terms in the 2PCF model.

\citet{Scoccimarro1999, GilMarin2015} have modeled 
 the effect of incoherent motions with a damping term in the galaxy bispectrum
\begin{equation}
\label{eq:damping_bispectrum}
D^{B}(\alpha\sigma_{12}) = \left\{1+\alpha^{2}\left[\left(k_{1} \nu_{1}\right)^{2}+\left(k_{2} \nu_{2}\right)^{2}+\left(k_{3} \nu_{3}\right)^{2}\right]^{2} \frac{\sigma_{12}^{2}}{2} \right\}^{-2}.
\end{equation}
The magnitude of the damping effect is determined by the pairwise velocity dispersion, $\sigma_{12}$ and modulated, in the model,  
by a multiplicative factor, $\alpha$ (with $\alpha=0$
describing a purely coherent flow).
In this work, however, we decided to drop this damping term since 
it has a negligible impact on scales larger than $ 10 \, \Mpch$ as we show in a dedicated test in Sect.~\ref{sec:3pt_results}, where we assess the sensitivity of our results to the choice of $s_{min}$.
This choice has the further advantage of avoiding computationally demanding
2-dimensional Fourier transforms.

\subsection{Integral constraint}

The so-called integral constraint effect originates from the assumption that the mean density estimated from the data coincides with the true one. This assumption biases the estimate of the correlation function on scales comparable to the size of the survey \citep{Hui99}, where it cannot be neglected, in particular at BAO scales  \citep{Slepian2017b}.
However, the maximum scales considered in our analysis (40 $\Mpch$) are significantly smaller than the size of the survey. This guarantees that the integral constraint is negligible, as shown by  \citet{Cappi2015}, and will be ignored in our analysis.

\subsection{Summary of model parameters}

Our 2PCF and 3PCF models are specified by the set of free parameters listed in
 Table~\ref{tab:parameters}.
These include: the clustering {\it rms} amplitude, $\sigma_8$, the linear redshift distortion parameter $\beta\equiv f/b_1$, the pairwise velocity dispersion, $\sigma_{12}$, and the two galaxy bias parameters, $b_1$ and $\gamma \equiv b_2/b_1$.

The other model parameters that are kept fixed to their fiducial values are: the shape parameters of $P_{lin}(k)$
which are set to the Planck values \citep{Planck18}; the VIPERS {\it rms} redshift error, $\sigma_{z}=5\times10^{-4}(1+z)$; the bias parameters $b_{s^2}$ and $b_{3nl}$, which are derived from local Lagrangian theory, i.e.
\begin{align}
    \label{eq:nlbias}
    b_{s^2} & = -\frac{4}{7} \left( b_1 -1 \right) \, , \\
    b_{3nl} & = \frac{32}{325} \left( b_1 -1 \right) \, .
\end{align}
This choice is motivated by the fact that their measured values would be too noisy, as we have verified by running a dedicated likelihood analysis, aimed at testing the local Lagrangian bias hypothesis. 
Finally, we set $\alpha=0$; a choice that has a 
negligible impact on the results, as we show in Sect.~\ref{sec:3pt_results}. 
\section{Parameter Inference}
\label{sec:posterior}

To estimate the free parameters of the model, described as a vector $\boldsymbol{\theta}$, and their uncertainties, we compute the posterior probability,  $P \left(\boldsymbol{\mu} (\boldsymbol{\theta}) | \vec{d} \right) $, where $\boldsymbol{\mu}(\boldsymbol{\theta})$ is the model prediction vector and $\vec{d}$ is the data vector. 

We perform three different analyses: one in which the data vector is represented  by the VIPERS 2PCF monopole and quadrupole measured in the two samples G1 and G2 in the range $[15,40] \, \Mpch$; one where 
we consider the measured 3PCF monopole of the same galaxies in the same range of scales; and one considering both statistics.
In each case, we use the corresponding numerical covariance matrix  $C$ as described in Sect.~\ref{sec:covmatrix}.

To evaluate the posterior probability, we multiply the likelihood, $\mathcal{L}(\vec{d}|\boldsymbol{\mu}, C)$,
by the prior probability, $P \left(\boldsymbol{\theta} \right)$.
For each free parameter, the prior is modelled as a step function
with zero value outside the ranges specified in Table~\ref{tab:parameters}.
To compute the likelihood $\mathcal{L}(\vec{d}|\boldsymbol{\mu}, C)$,
we use the covariance matrices 
described in Sect.~\ref{sec:covmatrix}. The latter have been numerically estimated from a large (153) but finite number of mock VIPERS catalogues, which introduces a systematic error. To correct for this, we adopt the approach of \cite{Sellentin2016} and use the modified likelihood function
\begin{equation}
    \mathcal{L} \left(\vec{d} | \boldsymbol{\mu}, C, N_m\right)= 
    \bar{c}_{\mathrm{p}} \left(\det{C}\right)^{-1/2}\left[1+\frac{\chi^2(\boldsymbol{\mu},C)}{N_m-1}\right]^{-\frac{N_m}{2}} \, ,
    \label{eq:lik_sellentin}
\end{equation}
where 
\begin{equation}
    \label{eq:chi2}
    \chi^2(\boldsymbol{\mu}, C) = \left(\vec{d} - \boldsymbol{\mu} \right)^T C ^ {-1}   \left(\vec{d} - \boldsymbol{\mu} \right) \, 
\end{equation}

\begin{equation}
    \bar{c}_{\mathrm{p}} = \frac{\Gamma\left(\frac{N_m}{2}\right)}{\pi \left(N_m-1\right)^{p/2} \Gamma \left( \frac{N_m -\ell}{2} \right)} \, ,
    \label{eq:lik_sellentin_norm}
\end{equation}
with $N_m$ being the number of mock catalogues, and $\Gamma$ the Gamma function.

To sample the posterior $P \left(\vec{\mu} (\vec{\theta}) | \vec{d} \right) $ we use a Monte Carlo Markov Chain (MCMC) approach that generates a set of chains, i.e. collections of points in the parameter space $\lbrace \vec{\theta}_1, \vec{\theta}_2, ..., \vec{\theta}_N \rbrace$. 
Some of the parameters used in our analysis are expected to be (partly) degenerate. In these cases, we do not consider the individual parameters, but a suitable combination, as e.g. for $b_1\sigma_8$ and $f\sigma_8$ in the case of the 2PCF analysis alone.

\begin{table*}
\caption{Priors for the free parameters used in our likelihood analyses. The pairwise velocity dispersion, $\sigma_{12}$ is expressed in $\Mpch$.}

{\renewcommand\arraystretch{1.2}
\begin{tabular}{|c|c|c|c|c|c|}
\hline
  \diagbox[width=10em]{Analysis}{Parameters}  & $b_1$  & $\gamma$  & $\beta$  & $\sigma_8(z)$ & $\sigma_{12}$ \\ \hline
2PCF  & [0.1, 5]  & [-5., 5]   & [0.1, 5] & [0, 5]  & [0, 10]  \\ \hline
3PCF  & [0.1, 5]  & [-5, 5]  & [0.1, 5] & [0.1, 5]  & -- \\   \hline
2PCF+3PCF & [0.1, 5]  & [-5, 5]  & [0.1, 5] & [0, 5]  & [0, 10] \\  \hline

\end{tabular}

\label{tab:parameters}
}
\end{table*}

\begin{figure*}
    \begin{tabular}{p{0.5\textwidth} p{0.5\textwidth}}
        \vspace{0pt} \includegraphics[width=0.49\textwidth]{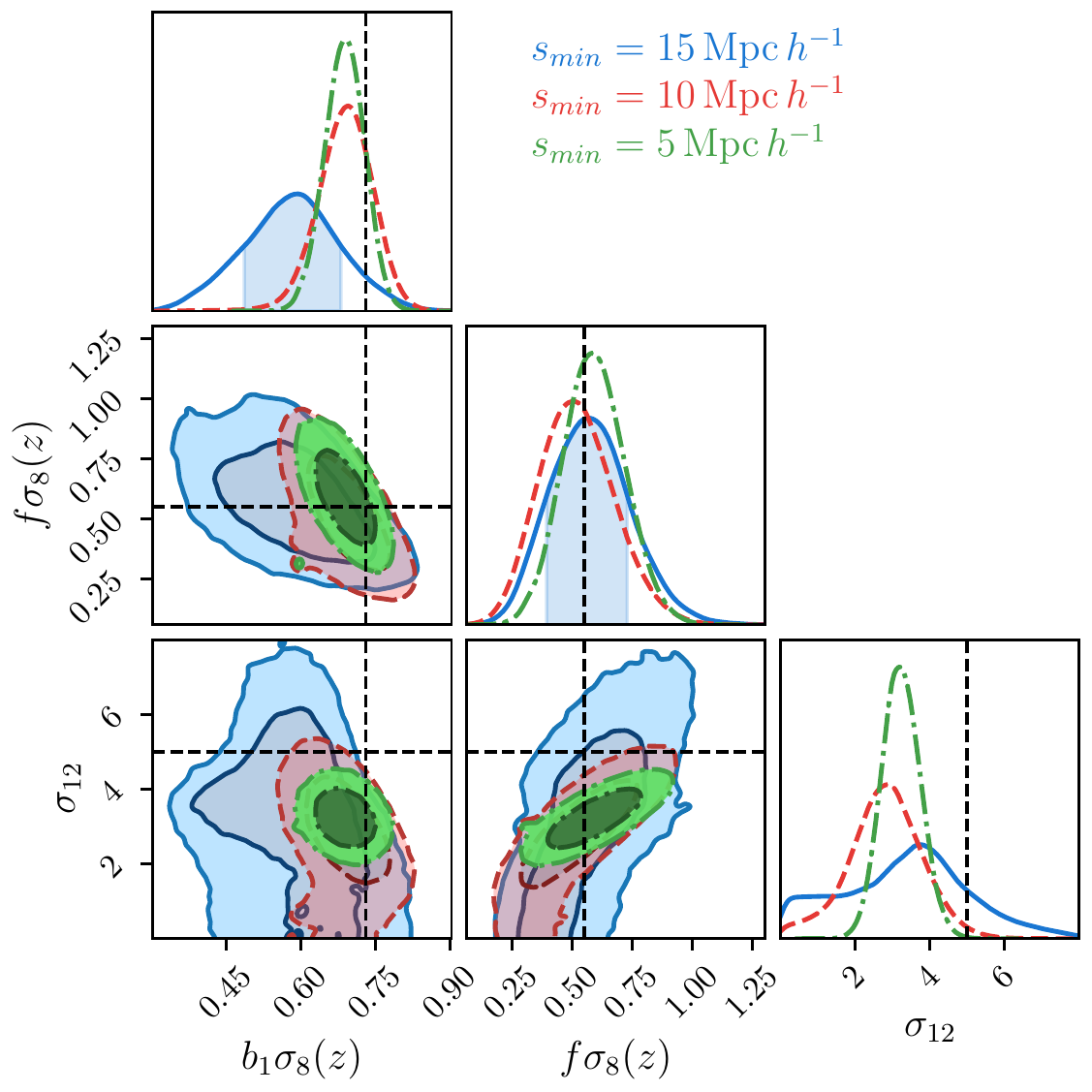} &
        \vspace{0pt} \includegraphics[width=0.49\textwidth]{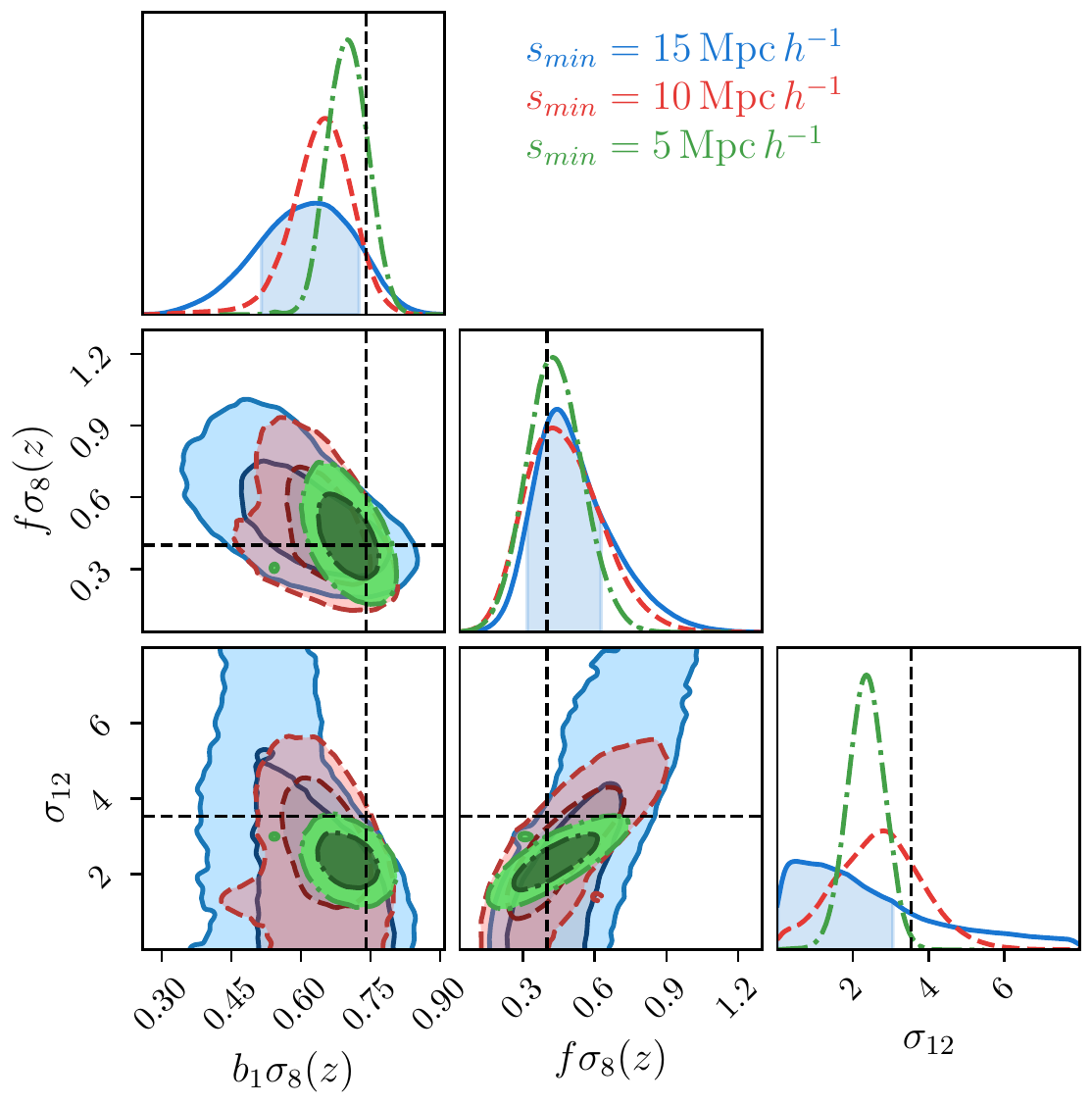}
\end{tabular}
\caption{
	Marginalized 2D posterior probability levels for the parameters $b_1 \sigma_8$, $f \sigma_8$ and $\sigma_{12}$, from the 2PCF analysis of the P1 and P2 ``control" samples (left/right respectively), as a function of the minimum scale included in the analysis (see legend). The units of $\sigma_{12}$ are $\Mpch$ and the
	darker/lighter shades for each set correspond to 68\% and 95\% confidence levels, respectively. 1D distribution functions on the diagonal give the 1D marginalized probability distributions of each parameter, with the 68\% interval indicated by the blue-shaded area for the $s_{min} = 15 \, \Mpch$ reference case only.  The horizontal and vertical dashed lines give the best-fit values obtained by \citet{Pezzotta2017} on similar sub-samples of the VIPERS PDR2 catalogue.
}
\label{fig:Pezzotta_comparison}

\end{figure*}

\begin{figure*}
\begin{tabular}{p{0.5\textwidth} p{0.5\textwidth}}
        \vspace{0pt} \includegraphics[width=0.49\textwidth]{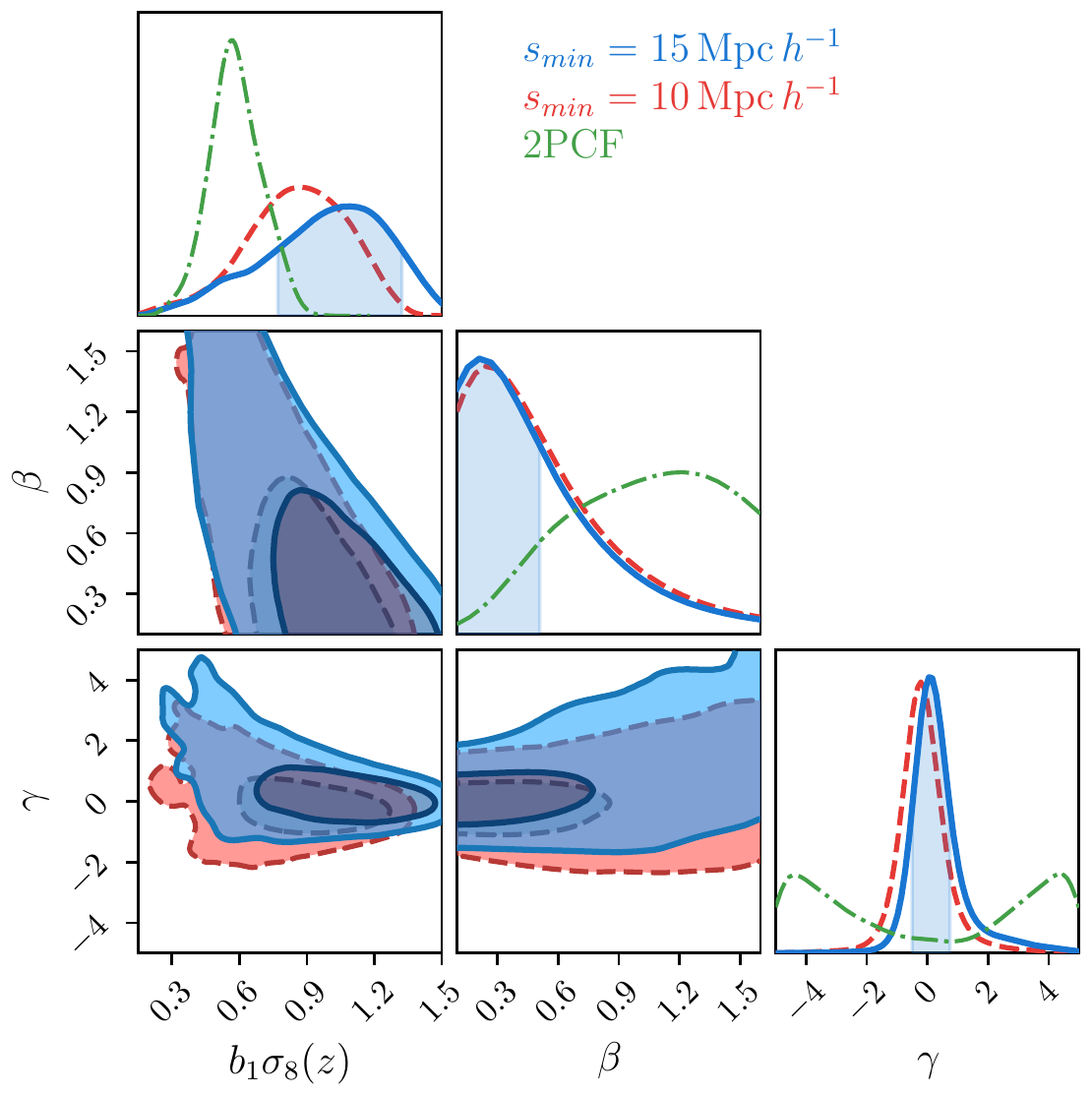} &
        \vspace{0pt} \includegraphics[width=0.49\textwidth]{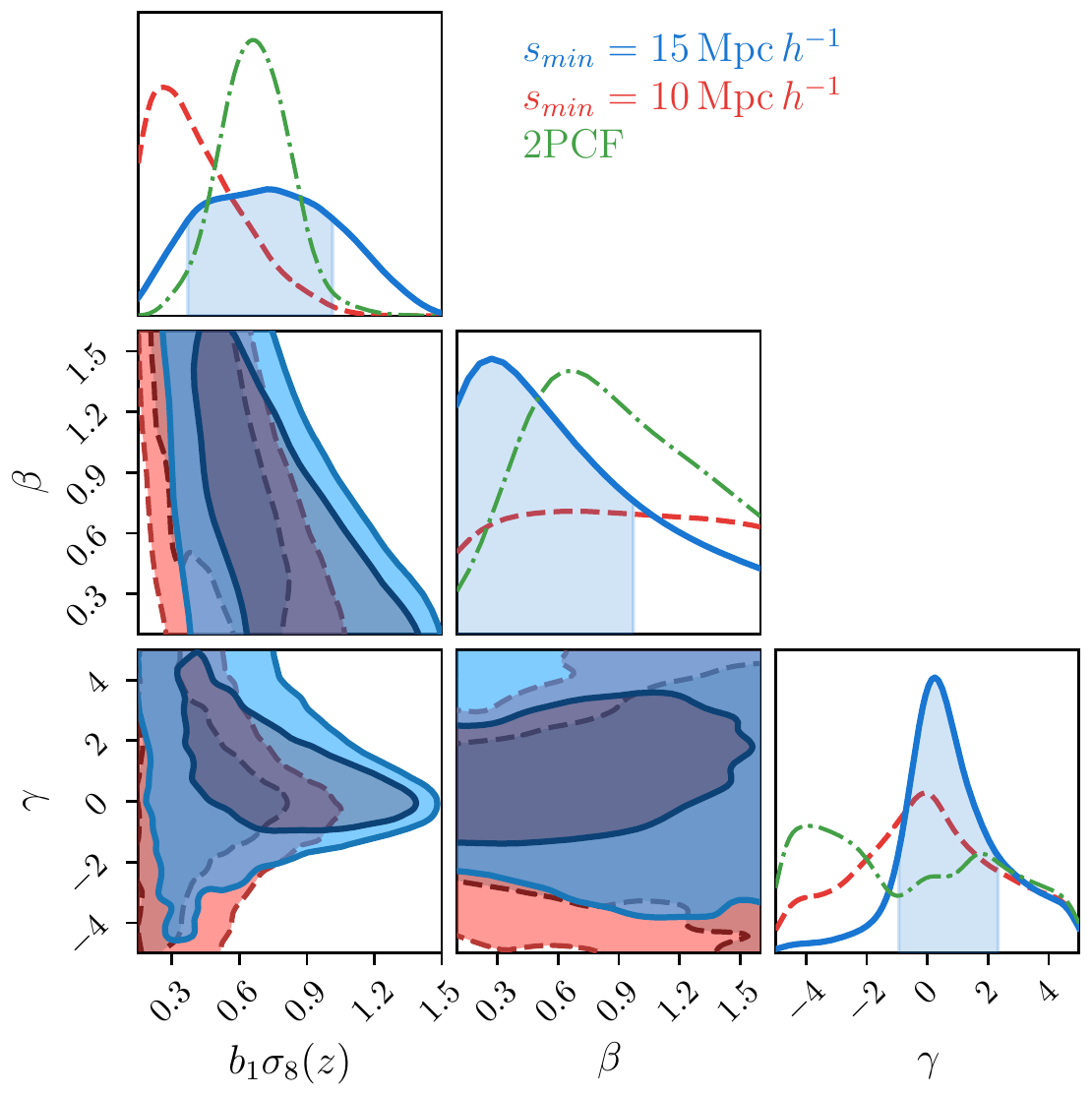}
\end{tabular}
	\caption{Marginalized 2D and 1D posterior distributions for the parameters $b_1 \sigma_8$, $\beta$ and $\gamma$, obtained from the 3PCF analysis of the G1 and G2 VIPERS samples (left and right panels, respectively). The plots show the results for the two choices of $s_{min}$ as indicated by the legend. As in previous figures, darker/lighter shades correspond to 68 \% and 95 \% confidence levels, respectively. The green dashed lines report the marginalized 1D distributions obtained for the same parameters from the 2PCF analysis, using $s_{min} = 15 \, \Mpch$.
	}
    \label{fig:3pcf}

\end{figure*}

\section{Results}
\label{sec:results}

In this section, we present the results of three analyses based, respectively, on the VIPERS
2PCF only, 3PCF only, and on the combination of the two statistics. 

The first analysis is very similar to those of \citet{DeLaTorre2013, DeLaTorre2017, Pezzotta2017, Mohammad2018} and we use it to validate our likelihood pipeline. 
The main goal of the second analysis, which builds upon the work of  \citet{Moresco2017},
is to measure the bias parameters $b_1$ and $\gamma\equiv b_2/b_1$.
Finally, the joint 2PCF and 3PCF analysis is aimed at breaking parameter degeneracies to obtain independent constraints on the galaxy bias, the evolution of cosmic structures and the clustering amplitude.

For all such analyses, the G1 and G2 VIPERS sub-samples are used (see Table~\ref{tab:samples}). For the 2PCF study only, we also consider the P1 and P2 samples to allow us a direct comparison to \citet{Pezzotta2017}.

\subsection{2PCF} 
\label{sec:2pt_results}

The results of the 2PCF analysis are shown in Fig.~\ref{fig:Pezzotta_comparison}. In the panels
we plot the 2D and 1D posterior probability distributions for the combinations of degenerate parameters $f \sigma_8$, $b_1 \sigma_8$ 
and for $\sigma_{12}$, obtained from the analysis of the VIPERS P1 ($\bar{z} = 0.61$) and P2 ($\bar{z} = 0.87$) samples and for different $s_{min}$ values indicated in the plots.  We  considered the 2PCF monopole and quadrupole moments and used the $10 \times 10$ covariance sub-matrix  $C_{\xi, \xi}$ of Fig.~\ref{fig:joint_corr}. 

Our analysis is very similar, though not identical, to that of \citet{Pezzotta2017}. To minimize the differences, we compare our results obtained by using an SPT 1-loop 2PCF model in redshift space with their ``TNS" model \citep{Taruya2010}.
For the same reason, we push the analysis down to $s_{min} = 5 \, \Mpch$, but then show the results obtained with larger values up to $s_{min} = 15 \, \Mpch$ (our reference case).

The 2D and 1D posterior distributions shown in Fig.~\ref{fig:joint_corr} agree well with the results  of \citet{Pezzotta2017} except for the pairwise velocity dispersion, $\sigma_{12}$, which is significantly ($2\sigma$) smaller. This is not surprising since, as shown in Fig. 18
of \citet{Pezzotta2017}, the \citet{Scoccimarro2004} 2PCF model tends to underestimate the value of $\sigma_{12}$.

The 2D contours in Fig.~\ref{fig:Pezzotta_comparison}
show a mild anti-correlation between $f \sigma_8$ and $b_1 \sigma_8$, which is expected since larger redshift distortions that boost up the 2PCF monopole can be compensated by reducing the clustering amplitude  $b_1 \sigma_8$. In addition, we find a positive correlation between $f \sigma_8$ and $\sigma_{12}$, which is explained by the fact that damping effects can be compensated by increasing the
clustering amplitude.

To assess the sensitivity  of our results to some of the model parameters,
we have performed two robustness tests.  In the first one we have repeated the likelihood analysis with
different values of $s_{min}= 5, \, 10 \, \mbox{and} \,  15 \, \Mpch$. 
The results are shown in the same Fig.~\ref{fig:Pezzotta_comparison}. As expected, increasing the value of $s_{min}$ amplifies the errors, especially for the nonlinear $\sigma_{12}$ parameter. However,
no systematic error is introduced as all results  agree with those of \citet{Pezzotta2017} and with each other.

As a second test we used the alternative 2PCF model introduced in Sect.~\ref{sec:2pointm}
in which we 1) use  {\small HALOFIT}  \citep{Takahashi2012} to model $P_{\delta \delta}$ and 2) adopt the fitting formula of  \citet{Bel2019} to model $P_{\delta \theta}$ and $P_{\theta \theta}$.
The results are very similar to those obtained with the reference 2PFC model.

Finally, we have repeated the likelihood analysis on G1 and G2 samples that we use for the 3PCF analysis too.
The results, which are qualitatively similar to those obtained with P1 and P2, are summarized in  Table~\ref{tab:model_params_fit} where we list the best-fit parameters and their $1\sigma$ uncertainties.

We confirm that the 2PCF analysis successfully constrains the parameter combinations $f \sigma_8$,  $b_1 \sigma_8$ and  $\beta \equiv f/b_1$ but leave $\gamma$ unconstrained. This is not surprising since this parameter is sensitive to the
small scale clustering that we ignore, having set $s_{min}=15 \, \Mpch$. Indeed, when the analysis is extended down to $s_{min}=5 \, \Mpch$, we do measure a $\gamma$  value significantly different from zero, yet mildly inconsistent with the results of previous VIPERS analyses \citep{Cappi2015, DiPorto2016}. This value is also in tension with the one yielded by the 3PCF analysis presented in the next section. It probably indicates the minimum scales below which our galaxy 2PCF model fails and thus
justifies our choice of setting $s_{min}=15 \, \Mpch$. 

\subsection{3PCF}
\label{sec:3pt_results}

We have repeated the likelihood analysis to compare the 3PCF measurements in the 
G1 and G2 sample with the 3PCF model described in Sect.~\ref{sec:3pointm} and the
$25 \times 25$ covariance matrix $C_{\zeta, \zeta}$ shown in Fig.~\ref{fig:joint_corr}.

This analysis improves upon the one performed by \citet{Moresco2017} in different ways. First, we use a more accurate 3PCF model and a covariance matrix estimated from realistic mock catalogues to perform a full likelihood analysis.
Secondly, here we use the final, PDR2 release of the VIPERS catalogue.
Finally, we consider all triangle configurations and not just a few subsets. We stress the importance of this latter aspect since, in our estimator, we use a triangle representation that is flexible enough to allow us to select subsets of triangles characterized by specific configurations and side lengths, greatly simplifying the comparison with model predictions. 

The plots in Fig.~\ref{fig:3pcf} are analogous to those of Fig. \ref{fig:Pezzotta_comparison}.
They show the 2D and 1D marginalized posterior distributions for the parameters $b_1\sigma_8$, $\beta$ and $\gamma$ obtained from the 3PCF measured in the G1 (left) and G2 (right) samples.
The best-fitting parameters and their uncertainties are listed in   Table \ref{tab:model_params_fit}.
The 3PCF analysis allows us to estimate the nonlinear bias parameter $\gamma$, which turns out to be consistent with zero, while  2PCF is 
completely insensitive to it, as shown by the green, dot-dashed curves in the bottom right panels. 
This result is in agreement with \citet{Moresco2017}
and marginally consistent with those obtained from counts-in-cells analyses of
\citet{Cappi2015} and \citet{DiPorto2016} where a non-zero value of $\gamma$ was measured but on scales much smaller than those considered here.

The constraints on the distortion parameter $\beta$ are, on the contrary, quite weak. The best-fit value of this parameter is consistent with zero within the errors.
This is not unexpected since in our analysis we considered only the 3PCF monopole moment, which is insensitive to the  RSD effects. Constraints on $\beta$ could be obtained by considering the 
3PCF multipoles. This is, however, 
beyond the scope of this work since a multipole analysis would require the use of a significantly larger covariance matrix, which cannot be accurately estimated with the 153 mock catalogues at our disposal.

Another interesting feature is the mild degeneracy between $\beta$ and the clustering amplitude $b_1\sigma_8$ also seen for the 2PCF. The difference is that here the error on $b_1\sigma_8$ from the 3PCF analysis is  about $3$ times larger than from the 2PCF one.

To test the sensitivity to the choice of $s_{min}$ we repeated the likelihood analysis
using $s_{min}=5, \, 10, \, 15 \, \mbox{and} \, 20 \, \Mpch$.
In Figure \ref{fig:3pcf} we only show the cases of $s_{min} = 10$ and  15 $\Mpch$ to avoid overcrowding.
When $s_{min}$ decreases, the maximum of the 
1D posterior distributions for  $b \sigma_8$
shifts to smaller values. However, due to the large errors, the best-fit values
of $b \sigma_8$ (but also the ones of $\gamma$ and $\beta$) are in fact consistent with each other for all choices of $s_{min}$. In Appendix \ref{sec:systematics} we present a more detailed study of the impact of $s_{min}$ on the parameters' estimates using the VIPERS mock catalogues.

Finally, despite their different shapes,  the 1D posteriors of all parameters obtained from the 3PCF analysis for all values of $s_{min}$ largely overlap those obtained from the 2PCF.

\begin{figure*}
	\includegraphics[width=\textwidth]{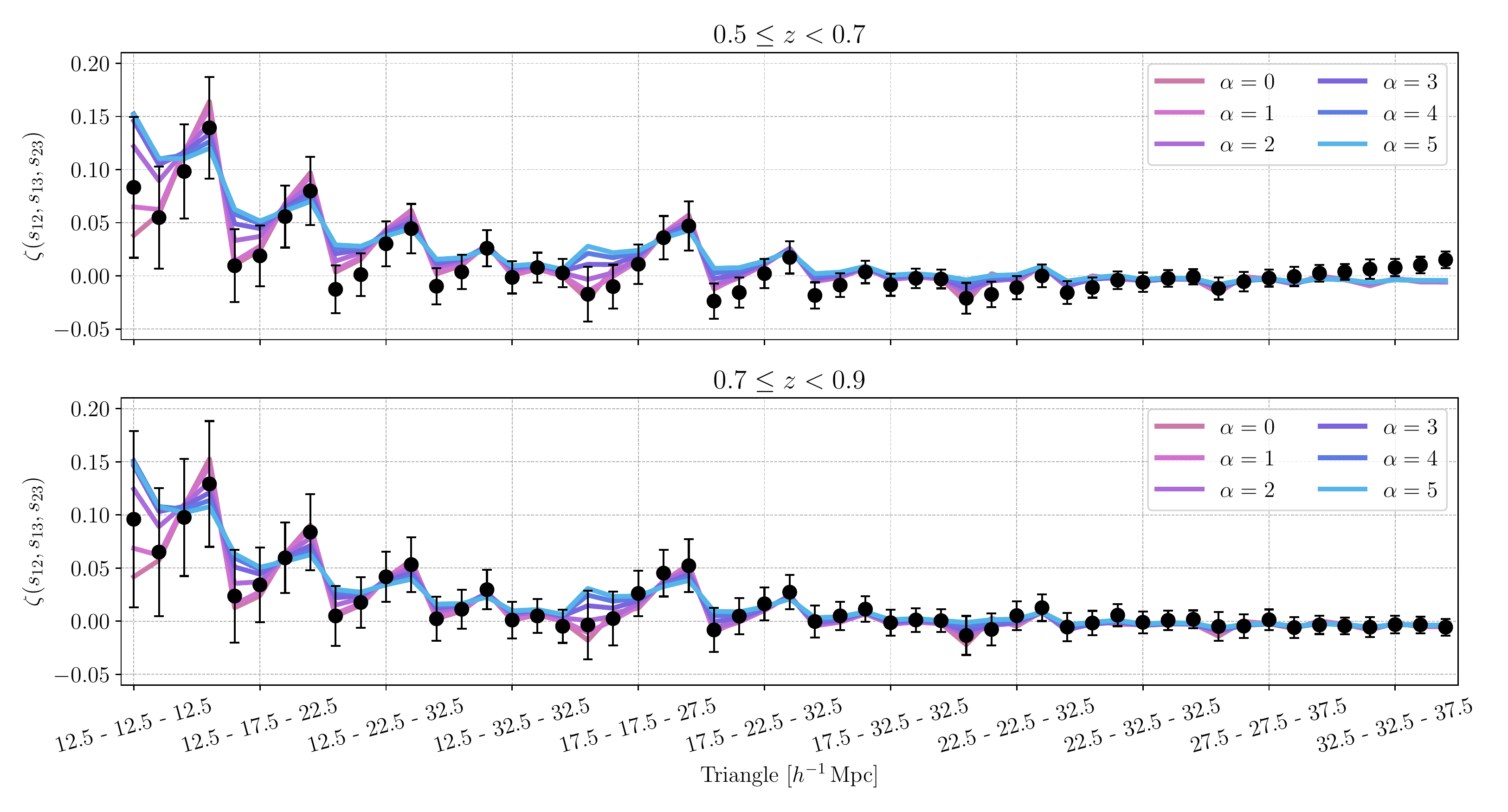}
	
	\caption{The impact of small-scale damping on our model for the 3PCF monopole.  The estimated 3PCF from the VIPERS G1 and G2 samples (top and bottom respectively, data and error bars as in Fig.~\ref{fig:3pcf_mono})
	are compared to the theoretical predictions, varying the amplitude of nonlinear motions, through the parameter $\alpha$. The case $\alpha=0 $ corresponds to the reference case, in which small-scales nonlinear motions are ignored.}
     \label{fig:model_non_linear_3pcf}
\end{figure*}

To further assess the impact of nonlinear effects, we compared the measured 3PCF with a model including the damping term of Eq. \eqref{eq:damping_bispectrum} and varying its strength.
The result is shown in Fig.~\ref{fig:model_non_linear_3pcf}, with the reference case with $\alpha=0$ corresponding to no damping.
We set the pairwise velocity dispersion, $\sigma_{12}$, equal to the best fit value obtained from the 2PCF analysis.
The plot compares model predictions with the 3PCF measured for various triangle shapes, whose side lengths are indicated on the x-axis. The results show that incoherent motions do not significantly affect model predictions on the scales considered in our analysis.
They may, however, become relevant for next-generation surveys, in which statistical errors will be significantly smaller than those considered here. We plan to further investigate this aspect and introduce a novel nonlinear 3PCF model to tackle the problem.

\subsection{Joint analysis of 2PCF and 3PCF}

\begin{figure*}
\begin{tabular}{p{0.5\textwidth} p{0.5\textwidth}}
        \vspace{0pt} \includegraphics[width=0.49\textwidth]{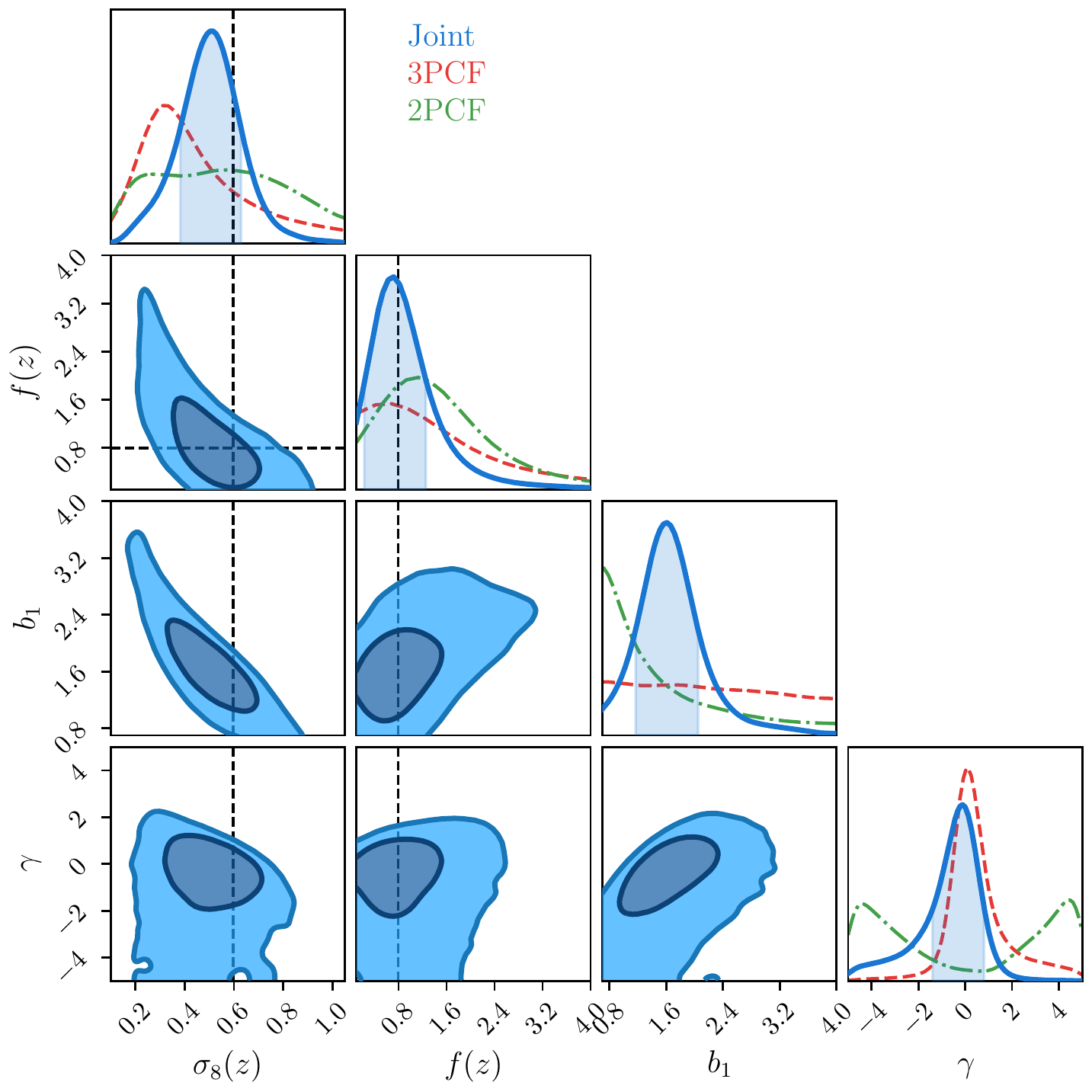} &
        \vspace{0pt} \includegraphics[width=0.49\textwidth]{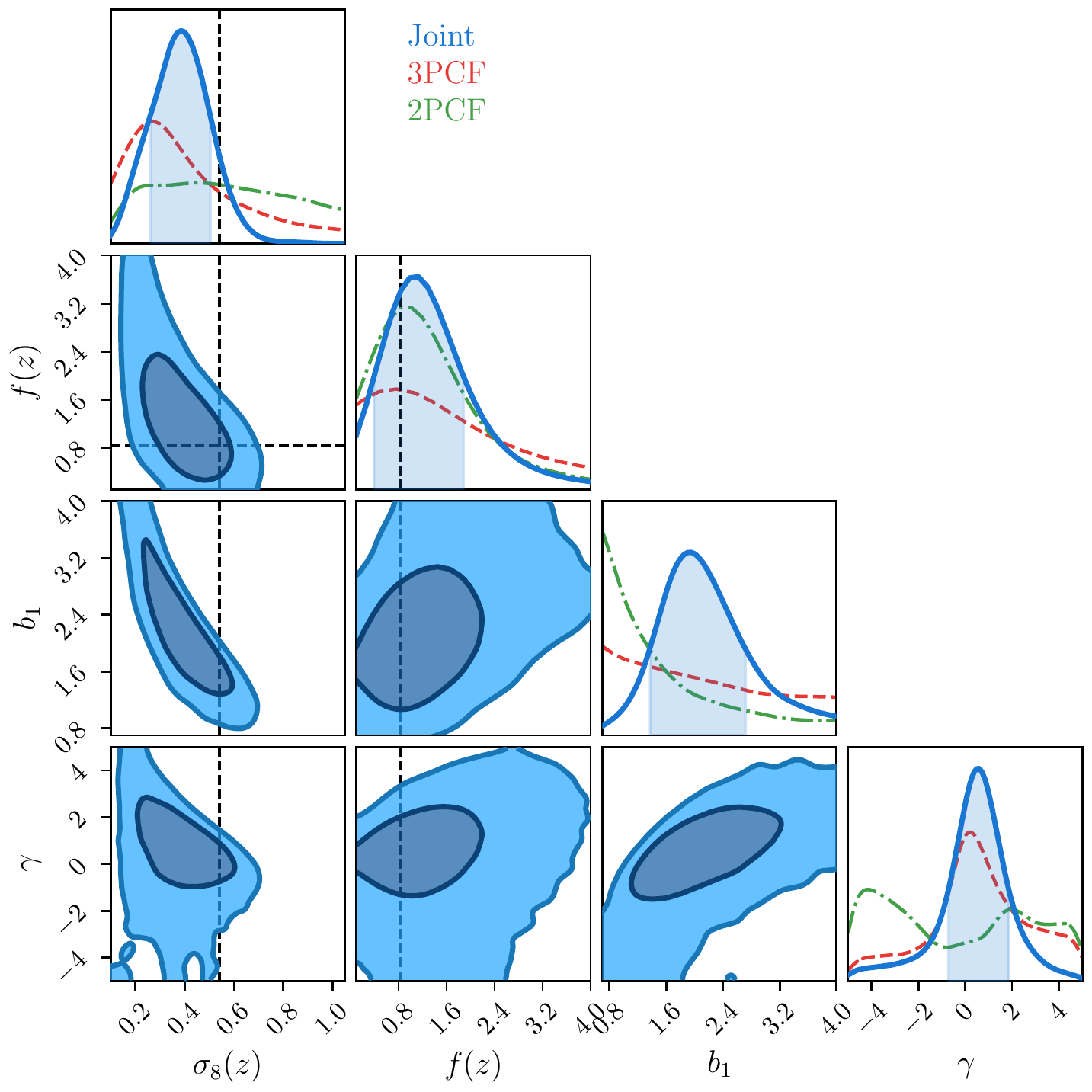}
    \end{tabular}
	\caption{Marginalized 1D and 2D posterior distributions for the parameters $\sigma_8(z)$, $f(z)$ and $b_1$, from the joint 2PCF+3PCF analysis of the G1 and G2 VIPERS samples (left and right panels, respectively). All estimates use $s_{min}=15 \, \Mpch$.  The meaning of contours and shaded areas are as in Fig.~\ref{fig:3pcf}.
	The vertical and horizontal black dashed lines show for reference the values measured for the same parameters by \citet{Planck18}, self-consistently scaled to the corresponding VIPERS redshift.}
	\label{fig:bias_cosmo_3pcf}

\end{figure*}

Finally, we perform for the first time a joint 2PCF and 3PCF correlation analysis on the VIPERS data. The main goal is to break the parameter degeneracy and obtain individual estimates for  $\sigma_8$, for the linear growth rate $f$, and for the two bias parameters $b_1$ and $\gamma$. 
 Fig.~\ref{fig:bias_cosmo_3pcf} shows the 1D and 2D posterior probability distributions (blue regions and solid curves) of these parameters from the likelihood analysis of the G1 and G2 samples and using the full 
 $45\times45$ covariance matrix.
The same range of scales $[15,40]  \, \mathrm{Mpc} \, h^{-1}$ was considered in both the 2- and 3-point statistics.
 The 1D posterior distributions obtained from the 2PCF-only (green, dot-dashed) and 3PCF-only (red, dashed) analyses are also shown for reference.
 The best-fit values of the joint analysis and their uncertainties are listed in  Table \ref{tab:model_params_fit}.
 
The joint analysis successfully breaks the $\sigma_8 - b_1 - f$ degeneracy, and the parameters can be measured individually, although with different uncertainties.
The value of $\sigma_8$  is measured with a $\sim 20 $ \% uncertainty in both G1 and G2 samples. The error on $b_1$ is $\sim 20$ \%  in the G1 sample, increasing to $\sim 30$\% in G2. The error on the growth rate $f$ is significantly larger ($65-75$\%), reflecting the fact that the $ f - \sigma_8 $ degeneracy is only partially broken by our analysis.
Though significantly less precise, it is reassuring that our best-fit $\sigma_8$ and  $f$ values are in agreement with those of the \citet{Planck18} cosmology,
scaled to the redshift of the survey (vertical and horizontal dashed lines).
The sensitivity of these results to the choice of $s_{min}$ is thoroughly discussed in Appendix~\ref{sec:systematics}.

The joint analysis does not improve the estimate of $\gamma$, whose precision is driven by the 3PCF signal, as shown in Fig.~ \ref{fig:bias_cosmo_3pcf}. It does instead improve the estimate of $\sigma_{12}$ whose value, in agreement with the one obtained from the 2PCF analysis, is estimated with a significantly higher precision.

Overall, the results of the joint analysis clearly show that the degeneracy of 
some parameters are successfully broken, thanks to the ability to extract information from intermediate to small scales where nonlinear effects are relevant, both in the evolution of the density fluctuation and in the biasing relation.

\begin{table*}
    \centering
    {\renewcommand\arraystretch{1.5} 
    \caption{Summary of the best-fit parameter values and their 1$\sigma$ uncertainties, as obtained from the G1/G2 VIPERS samples using the 2PCF-only, 3PCF-only and joint 2PCF+3PCF analyses.  The parameters common to all analyses are $b_1$, $\gamma$, $\beta$, $\sigma_8$. For the 2PCF and joint analyses, the pairwise velocity dispersion $\sigma_{12}$ is also considered. To account for parameter degeneracy, we also list some of the relevant parameter combinations constrained by the individual 2PCF and 3PCF analyses. }
    \begin{tabular}{cccccc|ccc|c}
        \hline
		Sample & Probe & $\sigma_8(z)$ & $b_1$ & $\gamma$ & $\beta$ & $\sigma_{12}$ &
		$f$ & $b_1 \sigma_8(z)$ & $f \sigma_8(z)$ \\ 
		\hline
		& 2PCF &  --  &  --  & - & $1.2\pm0.6$ & $2.4\pm2.1$ & -- & $0.56^{+0.12}_{-0.11}$ & $0.7\pm0.2$  \\ 		
		G1 & 3PCF & -- & -- & $0.1^{+0.8}_{-1.3}$ & $0.02^{+0.70}_{-0.0}$ & -- &
		--  & $1.0\pm0.3$ & $0.0\pm0.5$  \\ 
		& Joint & $0.50\pm 0.12$ & $1.60\pm0.43$ & $-0.1^{+0.8}_{-1.3}$ &
		$0.4^{+0.3}_{-0.2}$ & $2.4\pm 2.0$ &
		$0.64^{+0.55}_{-0.37}$ & $0.84^{+0.09}_{-0.14}$ & $0.36^{+0.17}_{-0.12}$  \\ 
		\hline
		 & 2PCF & -- & -- & -- & $0.6^{+0.9}_{-0.3}$ & $1.0^{+1.6}_{-1.0}$
		& -- & $0.67^{+0.15}_{-0.17}$ & $0.49^{+0.42}_{-0.16}$ \\ 
		G2 & 3PCF & -- & -- & $0.2^{+3.5}_{-1.3}$ & $0.0^{+1.4}_{-0.0}$ & -- &
		-- & $0.4^{+0.5}_{-0.2}$ & $0.0^{+0.7}_{-0.0}$ \\ 
		& Joint & $0.39^{+0.11}_{-0.13}$ & $1.9^{+0.8}_{-0.5}$ & $0.5^{+1.3}_{-1.2}$ &
		$0.49^{+0.31}_{-0.23}$ & $1.0^{+1.6}_{-1.0}$ &
		$1.0\pm 1.0$ & $0.74^{+0.07}_{-0.08}$ & $0.43^{+0.16}_{-0.15}$  \\ 
		
		\hline
    \end{tabular}
    }
    
    \label{tab:model_params_fit}
\end{table*}

\section{Discussion and Conclusions}
\label{sec:conclusions}

\begin{figure*}
\includegraphics[width=\textwidth]{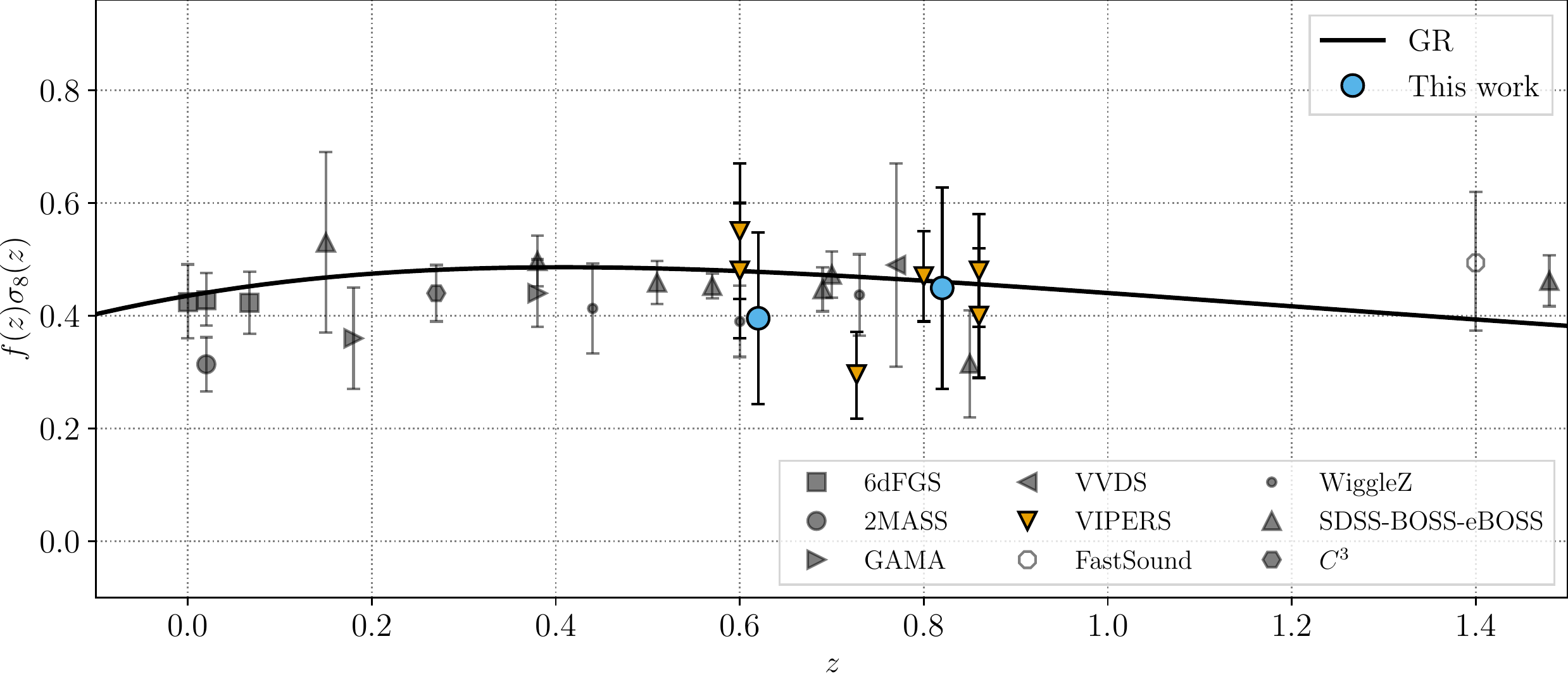}
\caption{Comparison of our new estimates of the growth rate of structure, parameterised by the product $f\sigma_8$ (light blue circles), with previous measurements from the literature:
6dFGS
    \citep{Beutler2012, Huterer2017, Adams2017}; 2MASS
    \citep{Davis2011}; GAMA \citep{Blake2013}; WiggleZ
    \citep{Blake2012}; VVDS \citep{Guzzo2008}; VIPERS
    \citep{DeLaTorre2013, DeLaTorre2017, Pezzotta2017, Hawken2017,
    Mohammad2018}; FastSound \citep{Okumura2016}; SDSS+BOSS+eBOSS
    \citep{eBOSS2020}, "$C^3$ - Cluster clustering cosmology" \citep{Marulli2020}. The black solid line shows the 
    $\Lambda$CDM + GR \citet{Planck18} model prediction.}
\label{fig:fs8}
\end{figure*}

\begin{figure*}
\includegraphics[width=\textwidth]{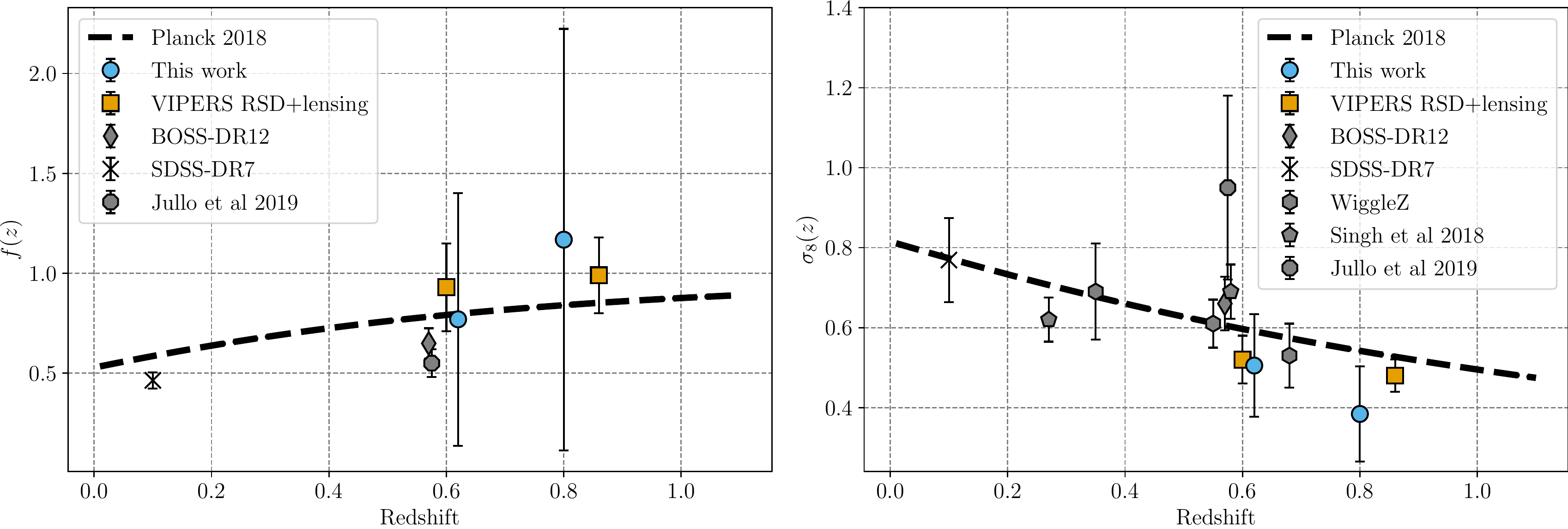}
\caption{ Values for $f(z)$ (left panel) and $\sigma_8(z)$, as estimated from our joint 2PCF and 3PCF analysis 
(light blue circles), again compared to other recent measurements using different techniques: VIPERS joint lensing-RSD analysis \citep{DeLaTorre2017},
     BOSS-DR12 \citep{GilMarin2017}, Wiggle-z \citep{Marin2013}. For completeness, we also include results at smaller redshifts by \citet{Shi2018}, \citet{Singh2019} and \citet{Jullo2019}.}
\label{fig:f_and_s8}
\end{figure*}

In this work, we have measured the 2PCF and 3PCF of the galaxies in the final data release of the VIPERS survey and used their joint information to break parameter degeneracies and estimate the galaxy bias parameters, $b_1$ and $\gamma$, the clustering amplitude, $\sigma_8$, and 
the linear growth rate of density fluctuations, $f$, at two redshifts $z\simeq 0.61$ and $z\simeq 0.8$.

Because of the survey footprint, separated in two patches highly elongated in one direction, we performed the analysis in configuration space focusing on scales smaller than $ 40 \, \Mpch $.
To do so, we adopted a 1-loop anisotropic 2PCF model to account for non-linear effects, galaxy bias and redshift distortions. For the 3PCF we considered the tree-level model of its monopole moment in redshift space proposed by \citet{Slepian2017a}.
To estimate the 3PCF we have used the SHD method of \citet{Slepian2015} whose efficiency allowed us   to measure the 3PCF of all 153 mock VIPERS catalogs and thus to estimate the covariance matrix used in the likelihood analysis.
Because of all these aspects, we believe that our analysis improves over the previous joint 2PCF and 3PCF analysis of the WiggleZ galaxy sample that was performed by 
\citet{Marin2013} using the brute-force  triplet counting estimator simplified analytic models for 2PCF, 3PCF and their covariance matrix.

The main results of this work can be summarised as follow:
\begin{itemize}
    \item The results of our 2PCF analysis agree, as expected, with those obtained by \citet{Pezzotta2017} in the range $[5,50] \, \Mpch $ 
    despite having used a different covariance matrix and a diffeent 2PCF model.
    Indeed, our estimates of
    \begin{align*}
        f\sigma_8(z=0.61) & =0.58\pm0.12 \nonumber \\
        f\sigma_8(z=0.87) & =0.43\pm 0.12 \nonumber 
    \end{align*}
    and 
    \begin{align*}
        b_1\sigma_8(z=0.61) & = 0.69\pm0.04\\
        b_1\sigma_8(z=0.87) & = 0.70\pm0.04
    \end{align*}
    match those  of \citet{Pezzotta2017} 
    ($f\sigma_8(z=0.61)=0.55\pm 0.12, \, f\sigma_8(z=0.87)=0.40 \pm 0.11$; 
    $b_1\sigma_8(z=0.61)=0.73\pm0.03, \, b_1\sigma_8(z=0.87)=0.74\pm0.04 $).

     When we consider pairs with $s_{min} \geq 10 $  $\Mpch$ the  1D posterior probability distribution of $\gamma\equiv b_2/b_1$ is flat, whereas it peaks at negative values when pairs at separations as small as 5   $\Mpch$ are included.
     Since a negative value is neither confirmed by previous VIPERS analyses \citep{Cappi2015,DiPorto2016} nor by our joint 2PCF and 3PCF analysis, we consider it as a piece of evidence that the 2PCF model is inadequate to describe galaxy clustering on such small scales. 
     Moreover, we found that including pairs with separations larger than 40 $\Mpch$, where shot noise starts to be significant, increases the computational cost of a less accurate covariance matrix and does not improve the quality of the fit.
     
     For all these reasons, we focused on a smaller scale range $[15,40] \, \Mpch $, and then assessed the robustness of our results to this choice.
     We then repeated the 2PCF analysis of G1 and G2 samples  using this range and found  
    $f\sigma_8(z=0.61)=0.47^{+0.17}_{-0.15}$. This is in good agreement with the estimates of \citet{DeLaTorre2017}, $f\sigma_8(z=0.6)=0.48\pm 0.11$, and of \citet{GilMarin2017},
    $f\sigma_8(z=0.57)=0.432\pm 0.022$.
    Our measurements are shown in Fig.~\ref{fig:fs8} together with existing estimates of $f\sigma_8$ obtained at different redshifts. As evident, there is an excellent agreement with previous results and with the theoretical expectation of a Planck cosmology \citep{Planck18}.
    
  \item We measured the 3PCF of the G1 and G2 samples in the same range as the 2PCF, $[15,40]$  $\Mpch$, for all triangle configurations. We compared  these measurements with the model predictions to estimate the bias parameters $b_1$ and $\gamma$ (for the remaining parameters we have the assumed a local Lagrangian bias model). We found $\gamma(z=0.61)=0.1^{+0.8}_{-1.3}$ and $\gamma(z=0.8)=0.2^{+3.5}_{-1.3}$, in tension with the result of \citet{Moresco2017} ($\gamma(z\simeq 0.6)=-0.47\pm 0.144$).
  The mismatch probably reflects the different scales probed by the two cases, since 
  \citet{Moresco2017} pushed their analysis to scales as small as $ 1 \Mpch$ (and fixed all parameters of the model except $b_1$ and $b_2$).

  \citet{DiPorto2016} estimated the biasing function from a count-in-cells analysis of the first data release of VIPERS. Among the shape parameters are used to characterize the biasing function in that analysis the one dubbed $B\equiv[1-\tilde{b}/\hat{b}]$, 
  can be directly compared to $\gamma$. The measured  $B$ values at the redshifts of the G1 and G2 samples and on the scale of $8$  $\Mpch$  ($B(z=0.6)=0.007\pm0.006$ and $B(z=0.8)=0.005\pm0.005$) agree with our estimates of $\gamma$,
  
  \citet{Cappi2015} also performed counts in cells to derive volume-averaged higher-order correlation functions from which they inferred the bias parameters $b_1$ and $b_2$.
  The scales considered in their analyses and the magnitude cuts are, however, different from ours. Nevertheless, it is reassuring that, similarly to our case, they did not detect significant deviations from a linear biasing. 
  
    \item We have shown that a joint  2- and 3-point correlation analysis of the VIPERS samples successfully breaks the degeneracy among  $\sigma_8$, $f$ and $b_1$ (and, consequently, $b_2$) and reduces the errors in the estimate of their combinations $f\sigma_8$, $b_1\sigma_8$ and $f/b_1$. It is instructive to compare our results to those of similar analyses, aimed at breaking parameter degeneracies by combining 2- and 3-point statistics in configuration \citep{Marin2013} and  Fourier space \citep{GilMarin2017} and by combining clustering and gravitational lensing measurements using the VIPERS PDR2 data \citep{DeLaTorre2017}.
    
    Our estimates of the clustering amplitude, $\sigma_8(z=0.61)=0.50\pm 0.12$ and $\sigma_8(z=0.8)=0.39^{+0.11}_{-0.13}$, are in good agreement with those obtained by \citet{DeLaTorre2017}($\sigma_8(z=0.6)=0.52\pm 0.06$  and $\sigma_8(z=0.86)=0.48\pm 0.04$), \citet{Marin2013} ($\sigma_8(z=0.55)=0.61^{+0.08}_{-0.09}$), \citet{GilMarin2017} ($\sigma_8(z=0.57)=0.66\pm 0.067$) and with the Planck $\Lambda$CDM predictions, as shown in the right panel of Fig.~\ref{fig:f_and_s8}.
    
    Similarly, our estimates of the growth rate $f(z=0.61)=0.64^{+0.55}_{-0.37}$ and $f(z=0.8)=1.0\pm1.0$ agree with those of \citet{DeLaTorre2017} ($f(z=0.6)=0.93\pm 0.22$  and $f(z=0.86)=0.99\pm 0.19$) and \citet{GilMarin2017} 
    ($f(z=0.57)=0.649\pm 0.076$) within the error that is much larger in our case. All these measurements are shown in the left panel of 
    Fig.~\ref{fig:f_and_s8} together with the Planck $\Lambda$CDM predictions \citep{Planck18}.
    
    We notice that the errors on both  $f$ and $\sigma_8$ in our case are significantly larger than in \citet{DeLaTorre2017}, despite having used similar datasets.  The reason for this is twofold. First of all, we have adopted a more conservative approach and considered $s_{min}=15$  $\Mpch$, whereas \citet{DeLaTorre2017} used $s_{min}=5$  $\Mpch$. This choice, motivated by the discrepant $\gamma$ value found in the 2PCF analysis, prevents us from accessing information encoded on smaller scales. If these are included, then the errors in the 
    measured $f$ and $\sigma_8$ significantly decrease towards values comparable to those of 
    \citet{DeLaTorre2017}, as illustrated in Appendix \ref{sec:systematics}.
    The second reason is that the clustering-lensing analysis is more efficient in breaking parameter degeneracy than the joint 2PCF-3PCF.  In fact, the latter combines the galaxy 2PCF, which has a $b_1\sigma_8$ degeneracy, to the galaxy 3PCF, which has a similar degeneracy, $b_1^3\sigma_8^4$.
    On the contrary, the lensing analysis has a  $b_1\sigma_8^2$ degeneracy, which is effectively broken when combined with the galaxy 2PCF.
    
    \item We have performed many tests to check the robustness of our results against non-linear effects, which are expected to be relevant on the scales probed here and are the reason for the conservative cut at $s_{min}=15$  $\Mpch$.
    Decreasing $s_{min}$ reduces parameters errors from the 2PCF only analysis but also picks up a negative value for $\gamma$, in mild tension (given the considerable uncertainty) with 
    the joint 2PCF-3PCF analysis and previous VIPERS analyses.
    On the contrary,  reducing $s_{min}$ does not have a significant impact on the 3PCF and the joint analyses. 
    In Table \ref{tab:joint_params} we list the best fit values of  $\sigma_8$, $f$, $b_1$ and $\gamma$ obtained for different choices of $s_{min}$. Errors increase when $s_{min}$ increases, as a result of the reduction of the number of pairs/triplets and scales included in the analysis. In particular, with $s_{min}=20 \, \Mpch$ no significant constraints can be set on $\gamma$.
    A detailed analysis of the sensitivity of our results to the choice of $s_{min}$ is presented in  Appendix \ref{sec:systematics}.
    
 \end{itemize}

\begin{table}
    \centering
    \caption{Best-fit values of the parameters obtained in the joint analysis of 2PCF and 3PCF, using different values of $s_{min}$. The values of $s_{min}$ are
    in $\Mpch$. All the constraints are compatible, within 1 $\sigma$.}
    {\renewcommand\arraystretch{1.5}
    \begin{tabular}{cccccc}
        \hline
		 Sample & $s_{min}$ & $\sigma_8(z)$ & $f$ & $b_1$ & $\gamma$ \\ 
		 \hline
            & $ 10$ & $0.4\pm0.1$ & $1.0^{+0.7}_{-0.5}$ & $1.8^{+0.4}_{-0.3}$ & $-0.1\pm0.6$ \\
         G1 & 15 & $0.5\pm0.1$ & $0.6^{+0.6}_{-0.4}$ & $1.6\pm0.4$ & $-0.1^{+0.9}_{-1.3}$ \\
            & 20 & $0.4^{+0.3}_{-0.2}$ & $1.1^{+1.6}_{-1.0}$ & $0.5^{+0.7}_{-0.4}$ & unconstrained \\
         \hline
            & 10 & $0.3\pm0.1$ & $1.4^{+1.2}_{-0.8}$ & $2.0^{+1.0}_{-0.6}$ & $-1.1^{+1.4}_{-2.2}$ \\
         G2 & 15 & $0.4\pm0.1$ & $1.0\pm1.0$ & $1.9^{+0.8}_{-0.5}$ & $0.5^{+1.3}_{-1.2}$ \\
            & 20 & $0.2^{+0.2}_{-0.1}$ & $0.9^{+1.4}_{-0.8}$ & $1.8^{+1.9}_{-0.9}$ & $-0.2^{+1.5}_{-3.9}$ \\
         \hline
		\hline
    \end{tabular} }
    \label{tab:joint_params}
\end{table}

Our work confirms the importance of  clustering analyses beyond 2-point statistics. Here, we were able to break parameter degeneracies using a relatively modest number of objects (23352 and 13046 for the G1 and G2 samples, respectively). In this respect, this work represents a successful pilot study in the preparation for the next generation spectroscopic surveys, such as the DESI project, the \textit{Euclid} \citep{EuclidRedbook} and the \textit{Roman} \citep{WFirst2019} space telescope missions. 
These surveys will be able to perform clustering analyses on scales much larger than those considered here. However, small scales will still encode an even larger amount of information.
Its extraction will require measuring higher-order statistics and comparing results to non-linear models, as we have done in this work.
In this respect, one of the main lessons learned from our analysis is the need to develop a full 1-loop model for the 3PCF, matching the 2PCF one, which we plan to present in future work.

The availability of such a model, along with that of a new generation of efficient 3PCF estimators \citep{Slepian2015,slepian2016}, would make higher-order clustering analyses in configuration space more palatable and a serious contender to more traditional Fourier-space methods. Bispectrum analyses enjoy the availability of fast estimators and non-linear models. However, they suffer from mode coupling induced by complex survey geometries, which is difficult to account for.
Comparing the performances of joint 2-point and 3-point clustering analyses in configuration and Fourier space for next-generation spectroscopic surveys is another open issue, which we plan to investigate in the future.

Finally, in this analysis, we were able to break parameter degeneracies using higher-order statistics. Alternatively, it can be broken by combining 2-point clustering and gravitational lensing, as in \citet{DeLaTorre2017}. 
Both approaches have advantages and disadvantages. The 2- and 3-point clustering analysis requires a single (spectroscopic) dataset to be performed. However, it is less efficient in removing parameters' degeneracy. The clustering-lensing analysis is more effective in breaking degeneracy but needs both a photometric and a spectroscopic survey to be performed. How to best combine these two types of analyses is another interesting issue that deserves a dedicated future study.

\section*{Data Availability}
The VIPERS PDR2 data, as well as the mock samples used here are publicly available from the VIPERS web site (\url{http://vipers.inaf.it}). The clustering measurements and covariance matrices from this paper are available from the authors, upon request.

\section*{Acknowledgements}
AV and EB thank Emiliano Sefusatti, Alexander Eggemeier, Elena Sarpa, Massimo Guidi for useful discussions.
This paper uses data from the VIMOS Public Extragalactic Redshift Survey (VIPERS). VIPERS
has been performed using the ESO Very Large Telescope, under the ``Large Programme" 182.A-0886.
The participating institutions and funding agencies are listed at \url{http://vipers.inaf.it}.
AV, EB, LG and LM are supported by ASI/INAF agreement n. 2018-23-HH.0
``Scientific activity for Euclid mission, Phase D" and INFN project ``InDark".
EB and LG are further supported by MIUR/PRIN 2017 ``From Darklight to Dark Matter:
understanding the galaxy-matter connection to measure the Universe".
EB is also supported by ASI/INAF agreement  n. 2017-14-H.O ``Unveiling Dark Matter and Missing Baryons in the high-energy sky". MM acknowledges the grants ASI n.I/023/12/0, ASI n.2018-23-HH.0, and support from MIUR, PRIN 2017 (grant 20179ZF5KS).
This research was supported by the Munich Institute for Astro- and Particle Physics (MIAPP)
which is funded by the Deutsche Forschungsgemeinschaft (DFG, German Research Foundation)
under Germany's Excellence Strategy- EXC - 2094 - 390783311.
LM also acknowledge the support from the
grant PRIN-MIUR 2017 WSCC32.




\bibliographystyle{mnras}
\nocite{*}
\bibliography{bibliography} 



\appendix

\section{3PCF estimator binning}
\label{app:3pcfest}

The algorithm of \citet{Slepian2015} used in this work estimates the multipole coefficients
of the 3PCF Legendre expansion in radial bins $\Delta_{r_{12}}$ and $\Delta_{r_{13}}$,
$\zeta_{l}(\Delta_{r_{12}}, \Delta_{r_{13}})$. To obtain an unbiased estimate of 
$\zeta(\Delta_{r_{12}}, \Delta_{r_{13}}, \Delta_{\mu})$ in bins $\Delta_{\mu}$, one needs to use bin-averaged Legendre polynomials $\BarLeg{l}{\Delta_{\mu}}$, i.e.
\begin{equation}
     \zeta (\Delta_{r_{12}}, \Delta_{r_{13}}, \Delta_{\mu}) \, =  \sum_{l=0}^{l_{max}} \zeta_{l}(\Delta_{r_{12}}, \Delta_{r_{13}})
     \BarLeg{l}{\Delta_{\mu}}\, ,
\end{equation}
where 
\begin{equation}
    \BarLeg{l}{\Delta_{\mu}}\equiv\BarLeg{l}{\mu_{min} \leq \mu \leq \mu_{max}} = \frac{1}{\Delta \mu}
    \left[ \frac{\Leg{l+1}{\mu}-\Leg{l-1}{\mu}}{2l+1} \right]_{\mu_{min}}^{\mu_{max}}\, ,
    \label{eq:legendre_angle_averaged}
\end{equation}
in the bin $\Delta_{\mu}=[\mu_{min}, \mu_{max}]$; when $l=0$,  $\BarLeg{l=0}{\Delta_{\mu}}=1$.

In this work, we express the 3PCF as a function of $r_{23}$, rather than $\mu$, with the two quantities related through the relation
\begin{equation}
    \mu = \frac{r^2_{12} + r^2_{13} - r^2_{23}}{2r_{12}r_{13}} \, ,
    \label{eq:triangle_relation}
\end{equation}{}
so that, given $r_{12}$ and $r_{13}$, the cosine angle $\mu$  varies in the range $[0,1]$, whereas $r_{23}$  varies between $|r_{12}-r_{13}|$ and $|r_{12}+r_{13}|$.
One can show that in this case the binned Legendre polynomials are of the form
\begin{equation}
    \BarLeg{l}{\Delta_{r_{12}}, \Delta_{r_{13}}, \Delta_{r_{23}}} \, =  \frac{32\pi}{V_{12} 
    V_{13}} \int k^2 I_l(k; \Delta_{r_{12}}) I_l(k; \Delta_{r_{13}}) I_0(k; \Delta_{r_{23}}) 
    \de k \, ,
    \label{eq:legendre_triangle_averaged}
\end{equation}
where $V_{12}$ and $V_{13}$ represent the volume of the spherical shells of width 
     $\Delta_{r_{12}}$, $\Delta_{r_{13}}$ respectively, and
     
 \begin{equation}
    I_l(k; \Delta_r) = \int_{r_{min}}^{r_{max}} r^2 j_l(k r) \de r \, ,
    \label{eq:spherical_bessel_integral}
\end{equation}
where $j_l(k r)$ are spherical Bessel functions.

\section{Explicit expressions for the 3PCF model}
\label{app:AlBl}

The explicit expression of the $A_l$ terms that appear in Eq. \eqref{eq:zeta_model_compact} is
\begin{equation}
\begin{aligned}
\label{eq:3pcfm_bias_terms}
    A_0 = & b_{1}^{3} \left\{ 
            \frac{34}{21} \left[ 1+\frac{4}{3} 
            \beta+\frac{1154}{1275} \beta^{2}+\frac{936}{2975} 
            \beta^{3}+\frac{21}{425} \beta^{4} \right] \right. \\
          & \left. +\gamma\left[1+\frac{2}{3} \beta+\frac{1}{9} 
            \beta^{2}\right] + \frac{16}{675}\beta^2 \gamma^{\prime} 
            \right\}, \\
    A_1 = & -b_{1}^{3}  
            \left[1+\frac{4}{3} \beta+\frac{82}{75} 
            \beta^{2}+\frac{12}{25} \beta^{3}+\frac{3}{35} 
            \beta^{4}\right],  \\
    A_2 = & b_{1}^{3}  \left\{ 
           \frac{8}{21}\left[1+\frac{4}{3} 
           \beta+\frac{52}{21} \beta^{2}+\frac{81}{49} 
           \beta^{3}+\frac{12}{35} \beta^{4}\right] \right. \\
          & \left. +\frac{32 \gamma}{945} \beta^{2} +
            \frac{5}{2}\left(\frac{8}{15}+\frac{16 \beta}{45}+\frac{344 
            \beta^{2}}{4725}\right) \gamma^{\prime} \right\}, \\
    A_3 = & -b_{1}^{3}   \left[\frac{8}{75} 
            \beta^{2}+\frac{16}{175} \beta^{3}+\frac{8}{315} 
            \beta^{4}\right], \\
    A_4 = & b_{1}^{3}  \left[-\frac{32}{3675}
            \beta^{2}+\frac{32}{8575} \beta^{3}+\frac{128}{11025} \right] .
\end{aligned}
\end{equation}

\noindent
In the same equation, the $f_l$ terms are:

\begin{equation}
    \begin{aligned}
    f_l(r_{12}, r_{13}, r_{23}) = 
    \left\{
        \begin{array}{lr} 
            \xi^{[l]}(r_{12}) \xi^{[l]}(r_{13}) 
            \mathcal{L}_l\left( \mu_{23} \right) + \text{cyc.} &  
            \text{ if $l$ is even}\\
            \\
            \left[ \xi^{[l+]}(r_{12}) \xi^{[l-]}(r_{13}) + \right. & \\ 
            \left. \xi^{[l+]}(r_{13}) \xi^{[l-]}(r_{12}) \right]
            \mathcal{L}_l \left( \mu_{23} \right) + \text{cyc.} & 
            \text{ if $l$ is odd}  \, , 
        \end{array}
    \right.
\end{aligned}
\end{equation}
where
\begin{equation}\begin{split}
\xi_{i}^{[l]} &=\int \frac{k^{2}\mathrm{d}k}{2 \pi^{2}} {2 \pi^{2}} P_{\rm lin}(k) j_{l}\left(k r_{i}\right) \\
\xi_{i}^{\left[l \pm \right]} &=\int \frac{k^{2}\mathrm{d}k}{2 \pi^{2}} k^{\pm 1} P_{\rm lin}(k) j_{l}\left(kr_{i}\right).
\end{split}\end{equation}
and $P_{\rm lin}(k) $ is the linear matter power spectrum.

\section{Sensitivity to nonlinear effects}
\label{sec:systematics}

\begin{figure}
    \includegraphics[width=0.5\textwidth]{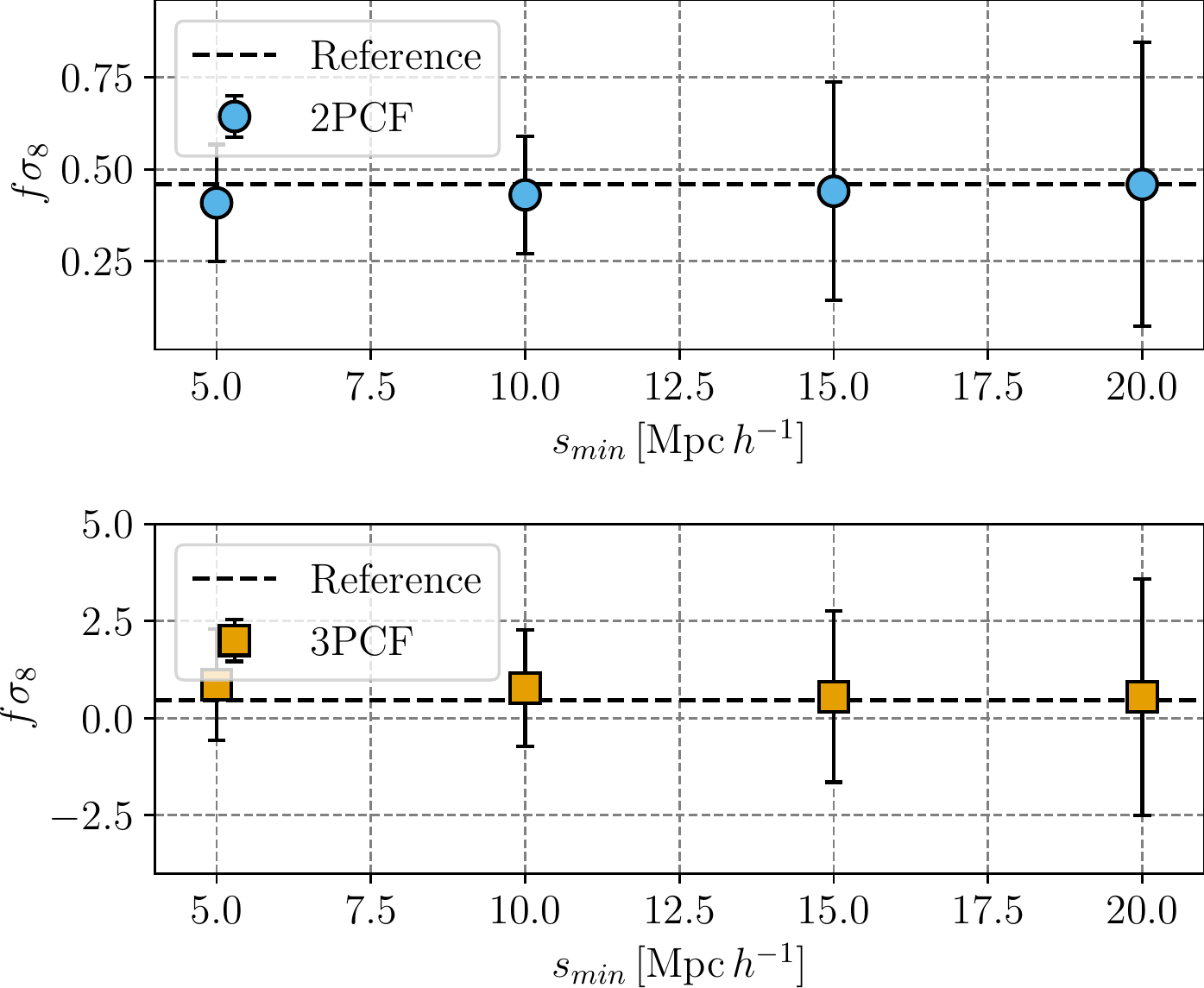}
    \caption{Average estimates over the 153 mock samples of $f\sigma_8$ using the 2PCF (top, filled points) and the 3PCF (bottom, filled squares), as a function of the minimum scale included in the analysis, $s_{min}$. The black dashed line in both cases gives the reference value, corresponding to the cosmology of the simulated mocks, while error bars are given by the scatter among the mocks. 
    Note how the error bars for the 3PCF estimates are $\sim 10$ times larger than those from the 2PCF, due to the fact that for the latter only the monopole (i.e. the isotropic information), has been considered.}
    \label{fig:3pcf_sys}
\end{figure}

\begin{figure}
    \centering
    \includegraphics[width=0.5\textwidth]{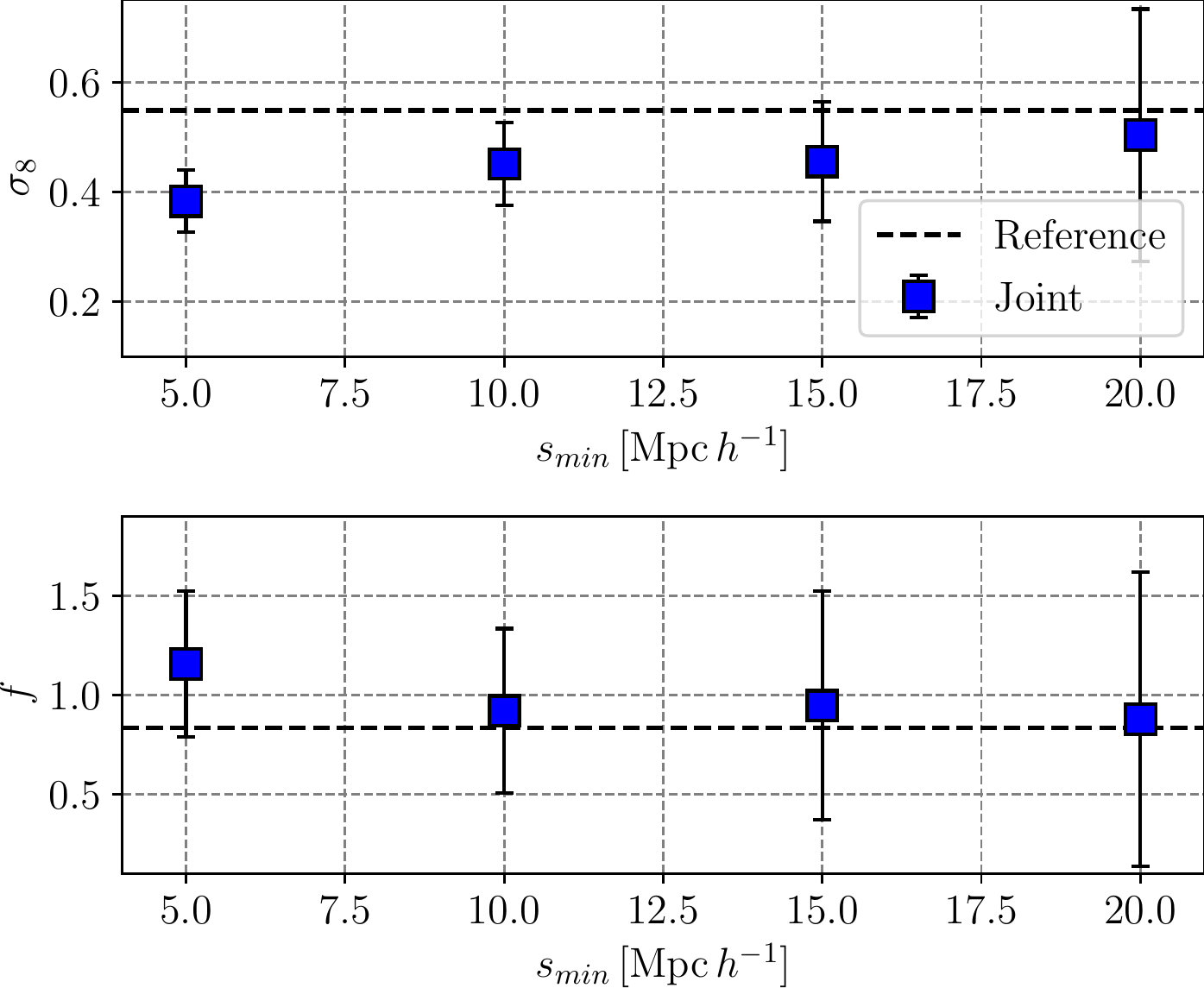}
    \caption{Average estimates of $f(z)$ and $\sigma_8(z)$ from the joint 2PCF-3PCF analysis of the 153 mocks (filled blue squares), compared to the reference cosmology (black dashed line). As usual, error bars are the standard deviation over the mocks. While the estimates of $f(z)$ are robust, recovering the input value of the mocks, there is a small bias for $\sigma_8$ (still within 1$\sigma$ at the reference $s_{min}=15 \Mpch$. }
    \label{fig:joint_sys}
\end{figure}

Since our clustering analysis includes scales smaller than $40 \Mpch$, nonlinear effects in both dynamics and galaxy bias cannot be neglected. 
To minimise the impact of potential systematic errors derived from 
an incorrect model of these effects, we have adopted a conservative approach and excluded scales below $15 \Mpch$ from the analysis. These are scales, however, where a significant amount of cosmological information is stored. In this Appendix, we 
use the realistic mock VIPERS catalogues to 
perform several tests to assess the sensitivity of our results to the choice of this minimum scale, $s_{min}$.
We are confident that these tests provide useful indications for the real data analysis since, as we have discussed in Sect.~\ref{sec:estimators}, the 2- and 3-point correlation functions measured in the mocks agree well with those  measured on the real data on all scales considered here.

Let us first consider the parameter combination $f \sigma_8$, as obtained from the 2PCF and the 3PCF. 
We show here the results from the G2 sample only, since they are representative of the generality of the results.
In Fig.~\ref{fig:3pcf_sys} we show the average and {\it rms} scatter over the 153 mocks of the value of $f \sigma_8$ estimated using the 2PCF and the 3PCF (top and bottom panels, respectively), as a function of $s_{min}$. 
The results show that including scales as small as $5 \, \Mpch$ significantly reduces the statistical error without compromising the accuracy. This conclusion is true for both the 2PCF and the 3PCF cases, indicating that the models we have used are adequate for measuring this parameter combination.

Things are different when one tries to break the parameter degeneracy through the joint 2PCF and 3PCF analysis, as shown in Fig.~\ref{fig:joint_sys}. In the two panels, the average estimates of $\sigma_8$ (top) and $f$ (bottom) are plotted as a function of $s_{min}$. Note how the size of the error bars decreases, when smaller scales are progressively included in the analysis. For $s_{min}= 5 \, \Mpch$ they are significantly smaller than in our baseline case $(s_{min}= 15 \Mpch)$. However, while random errors decrease, systematic errors increase: the value of $\sigma_8$ is systematically underestimated in general (although at the $1\sigma$ level only) and, to compensate, the measured growth rate is larger than the true one. 
These results justify our conservative choice to set  $s_{min}= 15 \, \Mpch$ and indicate that a better modelling 
is required to push a joint 2PCF and 3PCF analysis to smaller scales. 


\bsp	
\label{lastpage}
\end{document}